\documentclass[11pt,onecolumn,draftclsnofoot]{IEEEtran}
\usepackage{epsfig,graphics,subfigure,amsthm,psfrag,amsmath,amssymb}
\usepackage{algorithm} %//format of the algorithm
\usepackage{algorithmic} %//format of the algorithm
\usepackage{xcolor}
\usepackage{cite,citesort}
\newtheorem{Theorem}{Theorem}

\newtheorem{Definition}{Definition}

\newtheorem{Proof}{Proof}
\newtheorem{Proposition}{Proposition}
\begin{document}

\title{User Scheduling \\for Heterogeneous Multiuser MIMO Systems: \\ A Subspace Viewpoint}
\author{ \vspace*{0.25cm}
\begin{center}
\authorblockN{Xinping Yi and Edward Au, {\em Member}}
\thanks{X.~Yi is with the Mobile Communications Dept., EURECOM, 06560 Sophia Antipolis, France (email: yix@eurecom.fr)~and E.~Au is with Huawei Technologies, Shenzhen, China (email: edward.au@huawei.com).}
\end{center}
}
\maketitle %\vspace*{0.5cm} \thispagestyle{plain}\pagestyle{plain}

\begin{abstract}
In downlink multiuser multiple-input multiple-output (MU-MIMO) systems, users are practically heterogeneous in nature. However, most of the existing user scheduling algorithms are designed with an implicit assumption that the users are homogeneous.  In this paper, we revisit the problem by exploring the characteristics of heterogeneous users from a subspace point of view. With an objective of minimizing interference non-orthogonality among users, three new angular-based user scheduling criteria that can be applied in various user scheduling algorithms are proposed. While the first criterion is heuristically determined by identifying the incapability of largest principal angle to characterize the subspace correlation and hence the interference non-orthogonality between users, the second and third ones are derived by using, respectively, the sum rate capacity bounds with block diagonalization and the change in capacity by adding a new user into an existing user subset. Aiming at capturing fairness among heterogeneous users while maintaining multiuser diversity gain, two new hybrid user scheduling algorithms are also proposed whose computational complexities are only linearly proportional to the number of users. We show by simulations that the effectiveness of our proposed user scheduling criteria and algorithms with respect to those commonly used in homogeneous environment.
\end{abstract}

\begin{keywords}
Multiuser, MIMO,  Principal Angles, Subspace, User Scheduling.
\end{keywords}

\newpage
\section{Introduction}
Multiuser Multiple-Input Multiple-Output (MU-MIMO) systems have attracted a lot of interest and being considered as a promising technology in various beyond-3G standards~\cite{Li:10} due to its significant throughout improvement with respect to the single-user counterpart and its support of key features such as multiuser diversity and user multiplexing~\cite{Gesbert:07}. In order to achieve sum rate capacity of MIMO broadcast channels, one can apply Dirty Paper Coding~(DPC)~\cite{Costa:83,Jindal:05} but it comes at the expense of huge computational and implementation complexity.  Aiming at maximizing the sum rate capacity, a great number of low-complexity linear precoding algorithms have been proposed in which Block Diagonalization~(BD) is one of the popular choices due to its capability of approaching the capacity and its ease in practical implementation~\cite{Spencer:04-1}.

In an overloaded downlink MU-MIMO system that supports a very large number of users, user scheduling is necessary as base station~(BS) cannot usually serve such a large number of users simultaneously because of the following two key reasons.  First, there are far more users to be supported than the number of transmit antennas available at BS, which violates the dimensionality constraint of BD~\cite{Spencer:04-1}. Second, interference non-orthogonality among users always exists~\cite{Yoo:06}.  In other words, instantaneous channels among users are non-orthogonal to one another, which result in mutual inter-user interference.

There are generally two common types of scheduling algorithms, namely user selection and user grouping. For user selection algorithms whose objective is to select a subset of users for scheduling, it is natural to find an optimal subset by using exhaustive search but it is very computationally demanding even for moderate number of users. In this context, a large number of sub-optimal yet simplified algorithms has been proposed whose fundamental idea is to maximize system performance according to various user selection criteria~\cite{Dimic:05,Yoo:06,Bayesteh:08,Sigdel:09}. In~\cite{Yoo:06}, Yoo and Goldsmith construct a subset of semi-orthogonal users with a heuristic user selection criterion, which is shown to achieve an optimal capacity scaling when the number of users is asymptotically large.  Dimic~\emph{et al.}~\cite{Dimic:05}, propose a sub-optimal greedy algorithm to maximize the sum rate capacity by selecting a user who has the maximum projected norm on the null space of the existing users in a user subset. For users with multiple receive antennas, eigenmode and/or antenna selection is associated with user selection in order to achieve the best performance in terms of sum rate capacity~\cite{Bayesteh:08,Sigdel:09}. On the other hand, user grouping algorithms take fairness among users into account and schedule all users to be served over consecutive scheduling units, e.g.,~timeslots if Time Division Multiple Access (TDMA) is employed. In particular, all users are divided into a number of groups by certain criteria~\cite{Wang:06,Dawod:06}, whose aim is to maximize system performance while minimizing spatial correlation among users per group. Alternatively, one can consider applying proportional fair scheduling~(PFS)~\cite{Viswanath:02} into user selection algorithms for capturing fairness among users while maintaining multiuser diversity gain~\cite{Sigdel:09}. Though its performance is superior to the user grouping algorithms because of the multiuser diversity gain, PFS provides neither fairness nor delay guarantee~\cite{Yoo:06}.

For the above-mentioned user scheduling algorithms, an effective performance metric is required in selecting either an optimal subset of users or an optimal scheduling arrangement. There is a large body of literature focused on uncorrelated downlink MU-MIMO systems with homogeneous users~(see~\cite{Shen:06,Chen:08,Tran:10,Dao:10} and the references cited therein). For systems with heterogeneous users, the task of designing an efficient user scheduling metric becomes more challenging because there are more system parameters (for example, different number of receive antennas and different received signal-to-noise ratios~(SNRs)) to be considered. Though there are some recent works that consider scheduling strategies for users with different received SNRs~\cite{Sohn:10} and those with different number of receive antennas~\cite{Sun:10}, there is a lack of works addressing the combined problem based on the scheduling criteria. Naturally, an interesting question arouses in mind is, \emph{whether those heuristical scheduling criteria/metrics employed in homogeneous MIMO broadcast channels are still applicable in heterogeneous environment}. In this paper, we try to answer this question by studying users' channel characteristics in a subspace approach and designing effective user scheduling metrics from a geometric point of view. The main contributions of this paper are summarized as follows.
\begin{itemize}
\item
    In Section~III, we take into account the characteristics of principal angles between channels of heterogeneous users and propose three angular-based scheduling criteria that show superior performance than the existing ones in terms of sum rate capacity. These proposed criteria include geometrical angle (i.e.,~product of principal angles), grouping-oriented criterion that is derived by using an upper and a lower bounds of sum rate capacity with BD, and selection-oriented criterion that is derived by approximating the change in sum rate capacity due to the enrollment of a new user into an existing user subset.
\item
    In Section~IV, we propose two hybrid user scheduling algorithms that takes into account some key features of user grouping and selection algorithms, i.e., to capture fairness among users and to maximize the system performance in a greedy manner. When compared with the conventional user grouping algorithms~\cite{Wang:06,Dawod:06} that require an exhaustive search for all possible grouping arrangements, no brute-force search is required for our algorithms and their computational complexities are only linearly, rather than exponentially, proportional to the total number of users in the system. Further, simulation results in Section~V reveal the effectiveness of these algorithms in heterogeneous environment despite a reduction in the user's search space.
\end{itemize}

\textbf{Notation:}
Matrices and vectors are represented as uppercase and lowercase letters, respectively. Transpose and conjugate transpose of a matrix are denoted as~$(\cdot)^T$ and~$(\cdot)^H$, respectively. Further, we reserve~$diag \{ \cdot \}$ for an diagonal matrix, while~$\det(\cdot)$, $rank(\cdot)$, $tr(\cdot)$, $\Lambda(\cdot)$, $\lambda_i(\cdot)$, and~$|| \cdot ||_F$ represent the determinant, rank, trace, diagonal part, the $i$-th singular value, and Frobenius norm of a matrix.  In addition, we denote $\cap$ as an intersection of two subspaces, $(\cdot)^{\bot}$ as the null space of a subspace, and~$|\mathcal{\cdot}|$ as the cardinality of a set. Lastly, $E\{\cdot\}$, $\lfloor \cdot \rfloor$ and $abs(\cdot)$ denote the statistical expectation, floor operation, and absolute value operation, respectively.

\section{System Model}
Consider a downlink MU-MIMO system with $M_T$~transmit antennas at BS and $K$~heterogeneous users\footnote{Unlike homogeneous counterpart, heterogeneous users are of different antenna configurations and/or experience different channel environments.} that are equipped with $M_{R_k}$~receive antennas at the $k$-th user as illustrated in Fig.~1. We consider overloaded scenarios (i.e.,~$M_T \ll \sum_{k=1}^K M_{R_k}$) in which the BS cannot serve all the users simultaneously. For this reason, user scheduling is necessary to serve either a subset of users at one time or all users once over an entire scheduling period of $T$~timeslots.

Consider a subset of users~$\mathcal{T}$ that has been scheduled for transmission. Denote~$\mathbf x_k$ as the transmit signal of user-$k$ in the group (i.e.,~$k \in \mathcal{T}$). Its receive signal~$\mathbf y_k$ is given by
\begin{eqnarray}
    \mathbf y_k &=& \mathbf H_k \mathbf F_k \mathbf x_k + \sum_{l \in \mathcal{T}, l \ne k} \mathbf H_k \mathbf F_l \mathbf x_l + \mathbf n_k,
\end{eqnarray}
where~$\mathbf H_k \in \mathcal C^{M_{R_k}\times M_T}$ is the channel matrix between the BS and user-$k$ and it can be further expressed as $\mathbf H_k  = \sqrt{\rho_k} \bar{\mathbf H}_{k}$ with $\rho_k$~being the average received power of the $k$-th user and $\bar{\mathbf H}_{k}$ is an arbitrary matrix that depends on the channel model employed. Further, $\mathbf{F}_k$~is a precoding matrix of the $k$-th user. For practical consideration, we adopt BD~\cite{Spencer:04-1} as the linear precoder of the system. $\mathbf n_k$~is the additive white Gaussian noise~(AWGN) at the receive antennas of user-$k$ and is assumed to be zero-mean independent and identically distributed~(i.i.d.) Gaussian vector with variance~$\sigma_n^2$, i.e.,~$\mathbf n_k \sim \mathcal{CN}(0,\sigma_n^2)$.

Assume all channels are time-invariant during the scheduling period and the BS has perfect channel state information for all users. In order to track the influence of received SNR, we follow~\cite{Sohn:10} and consider it as a function of the distance between the BS and a user, path loss exponent, and average transmit power per antenna.

\section{User Scheduling Criteria for Heterogeneous Users}
In this section, we first briefly go over the existing user scheduling criteria for downlink correlated MU-MIMO systems with homogeneous users and then present our proposed scheduling criteria for heterogeneous users.

\subsection{Review on Metric Choice for Homogeneous Users}
Recall in Section~II that we consider an overloaded downlink MU-MIMO system in which there are a large number of users~$K$ and the total number of receive antennas is far greater than that of the total number of transmit antennas at the BS, i.e.,~$\sum_{k=1}^K M_{R_k} \gg M_T$. Since the number of users that the system can support is limited by the dimensionality constraint of BD~\cite{Spencer:04-1} and the instantaneous channel matrices among users are generally non-orthogonal, user scheduling is necessary.

Considering channel matrices as the subspace spanned by their column vectors, mutual interference across users can be represented as the correlation of the corresponding subspaces. Some recent works have been attempted to measure the level of subspace correlation in either angular or subspace domains and utilize them as user scheduling metrics. In the following, we will review three common ones, namely the largest principal angle, subspace collinearity, and chordal distance.

\subsubsection{Largest Principal Angle}
In order to facilitate our subsequent discussion, we first review the definition of principal angle, or equivalently, canonical angle~\cite{Afriat:57,Zhang:06}.
\begin{Definition}[Principal Angle~\cite{Bjorck:73,Absil:06}] For any two nonzero subspaces, namely,~$\mathcal{U}_k, \mathcal{V}_j \subseteq \mathcal{C}^n$ with~$p = \min \left\{ dim(\mathcal{U}_k), dim(\mathcal{V}_j) \right\}$, the principal angles between~$\mathcal{U}_{k,1} = \mathcal{U}_k$ and~$\mathcal{V}_{j,1} = \mathcal{V}_j$ are recursively defined to be the numbers~$0 \le \theta_{k,j,i} \le \pi/2$ such that
\begin{eqnarray}
    \cos \theta_{k,j,i} &=& {\max}_{\{\mathbf{u}_{k} \in \mathcal{U}_{k,i}, \mathbf{v}_{j} \in \mathcal V_{j,i}, \|\mathbf{u}_k\|_2 = \|\mathbf{v}_j\|_2 = 1\}} \mathbf{v}_j^H \mathbf{u}_k \\
    &=& \mathbf{v}_{j,i}^H \mathbf{u}_{k,i}, \quad i = 1,2,\cdots,p,
\end{eqnarray}
where~$\mathbf{u}_{k,i}$ and $\mathbf{v}_{j,i}$ are the vectors that construct the $i$-th principal angle $\theta_{k,j,i}$, $\|\mathbf{u}_{k,i}\|_2 = \|\mathbf{v}_{j,i}\|_2 = 1$, $\mathcal{U}_{k,i} = \mathcal{U}_{k,i-1} \cap \mathbf {u}_{k,i-1}^{\bot}$~and~$\mathcal{V}_{j,i} = \mathcal{V}_{j,i-1} \cap \mathbf{v}_{j,i-1}^{\bot}$~\footnote{From a space's viewpoint, $\mathcal{V}_{j,i-1} \cap \mathbf{v}_{j,i-1}^{\bot}$ represents the intersection of $\mathcal{V}_{j,i-1}$ and the null space of $\mathbf{v}_{j,i-1}$, which means that the representative subspace of $\mathbf{v}_{j,i-1}$ is removed from the subspace $\mathcal{V}_{j,i-1}$.}. Further,
\begin{eqnarray}
    \theta_{k,j,min} = \theta_{k,j,1} \le \theta_{k,j,2} \le \cdots \le \theta_{k,j,p} \le \theta_{k,j,max},
\end{eqnarray}
and $\theta_{k,j,p} = \theta_{k,j,max}$ when~$dim(\mathcal{U}_k)=dim(\mathcal{V}_j)$. $\hfill{\square}$
\end{Definition}

From the definition, we can find that the cosine of the principal angle is the inner product of two vectors of interest, and the minimal principal angle represents the largest inner product of any vectors in the two subspaces. If there exists intersection between two subspaces, the minimal principal angle would be zero.

In~\cite{Wang:06}, Wang and Murch proposed a user scheduling algorithm that aims at minimizing spatial correlation across users by arranging users with no or low inter-user spatial correlation into one group. In order to determine the impact of spatial correlation of the $k$-th user due to the remaining users, they utilize the largest principal angle between the orthogonal basis of the row spaces of user-$k$'s transmission channel~$\mathbf{H}_k$ and its interference channel~$\tilde{\mathbf H}_k = \begin{bmatrix} \mathbf{H}_1^T, \cdots, \mathbf{H}_{k-1}^T, \mathbf{H}_{k+1}^T, \cdots, \mathbf{H}_K^T \end{bmatrix}^T$. The resulting largest principal angle of each user is then used as a scheduling metric for optimal group arrangement. In Section~IV, we will present this algorithm in detail and discuss why it motivates us for proposing two reduced-complexity greedy-based hybrid user scheduling algorithms.

\subsubsection{Subspace Collinearity}
Subspace collinearity is a criterion that reflects the similarity of two matrix subspaces and it can be used for characterizing users' spatial separability~\cite{Czink:09,Ayach:09}.  In general, given two matrices~$\mathbf{M}_A$ and~$\mathbf{M}_B$, their collinearity can be represented as~\cite{Golub:96}
\begin{eqnarray}
    col(\mathbf{M}_A, \mathbf{M}_B) &=& \frac {abs\left(tr(\mathbf{M}_A \mathbf{M}_B^H)\right)} {\|\mathbf{M}_A\|_F \|\mathbf{M}_B\|_F},
\end{eqnarray}
where~$0 \le col(\mathbf{M}_A, \mathbf{M}_B) \le 1$, and~$col(\mathbf{M}_A, \mathbf{M}_B)=0$ when~$\mathbf{M}_A$ and~$\mathbf{M}_B$ are orthogonal to each other, while~$col(\mathbf{M}_A, \mathbf{M}_B)=1$ when the matrices are identical. It is clear that the smaller the collinearity is, the less similarity, or equivalently the larger distance, of the two matrix subspaces.

\subsubsection{Chordal Distance}
Chordal distance is commonly used in limited feedback systems~\cite{Love:03,Ravindran:08} for codebook design but it has also been recently considered as a user scheduling criterion~\cite{Ko:08}. As referred to~\cite{Conway:96}, the chordal distance between two subspaces $\mathcal U_k$ and $\mathcal V_j$ with dimensions $p$ and $q$ is defined in terms of the user's principal angles as follows
\begin{eqnarray} \nonumber
    d_c(\mathcal{U}_k, \mathcal{V}_j) = \frac{1}{\sqrt{2}} \| \mathbf P_{\mathcal{U}_k} - \mathbf P_{\mathcal{V}_j}\|_F = \sqrt{\sum_{i=1}^{\min\{p,q\}} \sin^2 \theta_{k,j,i}},
\end{eqnarray}
where $\mathbf P_{\mathcal{U}_k}$ and $\mathbf P_{\mathcal{V}_j}$ are the projection matrices of $\mathcal{U}_k$ and $\mathcal{V}_j$, respectively. Note that we only use the lower dimension, i.e.,~$\min\{p,q\}$, of the principal angles for the definition.

\subsection{Proposed User Scheduling Criteria}
In downlink MU-MIMO channel with heterogeneous users, principal angles between two subspaces of dimensions\footnote{Without loss of generality, we assume $p \le q$.}~$p$ and~$q$ possess the following specific characteristics, namely
\begin{eqnarray} \label{eq:ED-principal-angle-heter}
    0 = \underbrace{\theta_{k,j,1} = \cdots = \theta_{k,j,m}}_{Part~I} < \underbrace{\theta_{k,j,m+1} \le \cdots \le \theta_{k,j,n}}_{Part~II} < \underbrace{ \theta_{k,j,n+1} = \cdots = \theta_{k,j,p}}_{Part~III} = \frac{\pi}{2},
\end{eqnarray}
where~$\theta_{k,j,1}$ and~$\theta_{k,j,p}$ are the minimum and maximum principal angles, respectively. These three parts represent different physical meanings on subspace correlation (that models the mutual interference among users) and they are summarized as follows. For Part~I that consists of $m$~zero principal angles, they represent $m$~overlapped and fully-correlated basis of the two subspaces. As for the second part that is composed of $n-m$~principal angles whose values lie between~0 and~$\pi/2$, the corresponding basis of the two subspaces are non-overlapped but non-orthogonal with one another, which result in partial subspace correlation. Regarding the third part that contains $q-n$~principal angles of~$\pi/2$, it means there are $q-n$~orthogonal principal angles.

In order to understand the characteristics of~(\ref{eq:ED-principal-angle-heter}), we consider an example in which there are two heterogeneous users~$k$ and~$j$ with~$p = M_{R_k}$ and~$q = M_{R_j}$ receive antennas, respectively, and their instantaneous channel matrices are non-orthogonal to one another because of, e.g., their close proximity. Denote~$\mathcal{U}_k$ and~$\mathcal{V}_j$ as the two subspaces of dimensions~$p$ and~$q$ that are spanned by the columns of their channel matrices.  Assuming that the overlapped dimension of the channel subspaces is~$m$, we have~$dim(\mathcal{U}_k \cap \mathcal{V}_j)=m$ and hence $m$~zero principal angles (c.f.,~Part~I in~(\ref{eq:ED-principal-angle-heter})). For the non-overlapped counterpart of dimension~$p - m$, there exists $p - n$~mutually orthogonal components that correspond to $p - n$~largest principal angle of~$\pi/2$ (c.f.,~Part~III), while the remaining~$n - m$ ones are non-orthogonal with one another and they represent those principal angles with values between~0 and~$\pi/2$ (c.f.,~Part~II).

As referred to the characteristics above, it is clear that subspace correlation is reduced if the principal angles are as large as possible. For user scheduling algorithms that aim at minimizing subspace correlation and hence interference non-orthogonality among users, it means that the corresponding user selection/grouping criteria should be designed in such a way that users with a smaller dimension of Part~I and a larger dimension of Part~III can be served simultaneously.  This important observation leads us to understand that single-dimensional information, e.g.,~the largest/smallest principal angles, is sometimes not enough to characterize the correlation between two subspaces. For example, if~$\mathcal{U}_k$ and~$\mathcal{V}_j$ are subspaces of unequal dimensions that have a nontrivial intersection, then~$\theta_{k,j,min}=0$ and~$\theta_{k,j,max}=\pi/2$, but neither of them might convey the desired information on whether these two subspaces are highly correlated or not.  Similarly, though subspace collinearity reflects the similarity of two subspaces to some extent, it is an indirect measure because of its heuristic reflection of the orthogonality of the channel matrices. As for chordal distance,  it is similar to the subspace collinearity that requires the two compared subspaces $\mathcal U_k$~and $\mathcal V_k$~to be of the same dimension. If the subspaces are of different dimensions, the lower dimension is usually adopted, which may result in an inaccurate measure for systems with heterogeneous users.

In view of the disadvantages of these three metrics that do not accurately reflect users' spatial separability, it would be interesting to consider not only the largest and smallest principal angles, but also those ``intermediate angles''. In the following, we propose three other user scheduling criteria that take into account all principal angles as given in~(\ref{eq:ED-principal-angle-heter}).

\subsubsection{Geometrical Angle}
Let $\mathcal{U}_k = span\{\mathbf u_{k,1}, \mathbf u_{k,2},\cdots, \mathbf u_{k,p}\}$ and $\mathcal{V}_j = span\{\mathbf v_{j,1}, \mathbf v_{j,2}, \cdots, \mathbf v_{j,q}\}$ be two subspaces with $1 \le p \le q$. Geometrical angle, i.e., the angle~$\psi_{k,j}=\measuredangle(\mathcal{U}_k, \mathcal{V}_j)$ between the two subspaces, is defined as~\cite{Risteski:01,Gunawan:05}
\begin{eqnarray} \label{eq:user-grouping-geomatrical-angle}
    \cos^2 \psi_{k,j} &=& \prod_{i=1}^p \cos^2 \theta_{k,j,i},
\end{eqnarray}
where $\mathbf u_{k,i}$, $\mathbf v_{j,i}$ and $\theta_{k,j,i}$ are defined in Definition~1. Given two users~$k$ and~$j$ with channel matrices~$\mathbf H_k$ and~$\mathbf H_j$, respectively, without loss of generality, we assume~$p=rank(\mathbf H_k) \le rank(\mathbf H_j)=q$. Geometrical angle can be alternatively defined as~\cite{Gunawan:05}
\begin{eqnarray} \label{eq:user-grouping-geomatrical-angle-alter}
    \cos^2 \psi_{k,j} &=& \frac {\det(\mathbf M_{k,j} \mathbf M_{k,j}^H)} {\det(\mathbf H_k \mathbf H_k^H)},
\end{eqnarray}
where $\mathbf{M}_{k,j} = \mathbf{H}_k \mathbf H_j^H$~is a cross-correlation matrix that is represented in inner product form.

In general, the value of~$\cos^2\psi_{k,j}$ represents the ratio between the volume of the parallelepiped spanned by the projection of the basis vectors of the lower dimension subspace on the higher dimension subspace and the volume of the parallelepiped spanned by the basis vectors of the lower dimension subspace. Simply put, geometrical angle reflects the relationship between the projection of the smaller dimensional subspace onto the larger dimensional subspace and itself~\cite{Gunawan:05}. Fig.~2(a) shows a two-dimensional example.  By definition, geometrical angle refers to the ratio of the area of the projection of plane-A onto plane-B to that of plane-A itself. It is obvious that the ratio~$\cos^2\psi_{k,j}$ is larger (or equivalently, the angle~$\psi_{k,j}$ is smaller) when the planes are closer, and vice versa.

In our case, the level of subspace correlation between two users is characterized by the degree of overlapping between the corresponding channel subspaces. In particular, if the correlation is severe, the corresponding channel subspaces get closer to each other and hence the geometrical angle is smaller. In other words, there are more principal angles of zeros and less principal angles of~$\pi/2$ in~(\ref{eq:ED-principal-angle-heter}).  Aiming at minimizing subspace correlation and hence interference non-orthogonality among the scheduled users, an effective user scheduling criterion should select for users with larger geometrical angles~$\psi_{k,j}$ or equivalently, a smaller value of~$\cos^2\psi_{k,j}$, i.e.,
\begin{eqnarray} \label{eq:metric-geometrical-angle}
    \mathcal{M}(\mathbf H_k, {\mathbf H}_j) &=& \arg \min_{k,j \in \mathcal{C}} \cos^2 \psi_{k,j},
\end{eqnarray}
where $\mathcal{C}$~is the candidate user pool. The metric $\mathcal M(\mathbf H_k, \mathbf H_j)$ in~(\ref{eq:metric-geometrical-angle}) is introduced to denote the correlation of two channel matrices $\mathbf H_k$ and $\mathbf H_j$, and is used as a scheduling criterion for user grouping.

\subsubsection{Grouping-Oriented Criterion}
Recall from our discussion in~(\ref{eq:ED-principal-angle-heter}) that users with a smaller dimension of correlated basis~(i.e., Part~I) and a larger dimension of orthogonal basis~(i.e., Part~III) are preferred to be served together. Therefore, it is important for a user grouping algorithm to take into account subspace correlation among users so as to minimize interference orthogonality on each group while maximizing the sum rate capacity.  In the following, we re-express the sum rate capacity with BD in terms of principal angles and derive a scheduling metric that satisfies these two objectives.

\begin{Theorem}[Sum rate capacity bounds of BD]
For a $K$-heterogeneous user downlink MU-MIMO system, the sum rate capacity with BD is bounded as
\begin{eqnarray} \nonumber
    \sum_{k=1}^K \log_2 \left( 1 + \frac{\rho_k}{\sigma_n^2} \lambda_{k,min}^2 \sum_{i=1}^{M_{R_k}} \sin^2 \theta_{k,\tilde{k},i} \right) &\le& {C}_{sum}  \\ \nonumber
    &\le& \sum_{k=1}^K M_{R_k} \log_2 \left( 1 + \frac {\rho_k }{M_{R_k}\sigma_n^2} \lambda_{k,max}^2 \sum_{i=1}^{M_{R_k}} \sin^2 \theta_{k,\tilde{k},i} \right), \\ \label{eq:Theorem-1}
\end{eqnarray}
where~$\theta_{k,\tilde{k},i}$ is the $i$-th principal angle between the range space of the $k$-th user's channel~$\bar{\mathbf{H}}_k$ and its interference channel~$\tilde{\mathbf{H}}_k$. Moreover, $\lambda_{k,min}$ and~$\lambda_{k,max}$ are the minimum and maximum singular values of~$\bar{\mathbf H}_k$. $\hfill{\square}$
\end{Theorem}
\begin{Proof}
Please refer to Appendix~A for details.
\end{Proof}

It can be clearly observed from Theorem~1 that the sum rate capacity of the $k$-th user is increased monotonically with its~$\sum_{i=1}^{M_{R_k}} \sin^2 \theta_{k,\tilde{k},i}$ and the performance metric can be written as
\begin{eqnarray} \label{eq:metric-sum-sine}
    \mathcal{M}(\mathbf H_k, \tilde{\mathbf H}_k) &=& \arg \max_{k \in \mathcal{C}} \sum_{i=1}^{M_{R_k}} \sin^2 \theta_{k,\tilde{k},i}.
\end{eqnarray}
This metric reflects the correlation of channel matrices between user-$k$ and the other users that intend to group together, and it is used as a scheduling criterion for user grouping. In order to take into account those zero principal angles corresponding to the overlapped subspaces (c.f., Part~I in (\ref{eq:ED-principal-angle-heter})), however, we prefer to use the cosine function because the sine of zero principal angle is equal to zero. Hence, our grouping-oriented scheduling criterion is embodied as
\begin{eqnarray} \label{eq:metric-sum-cosine}
    \mathcal{M}(\mathbf H_k, \tilde{\mathbf H}_k) &=& \arg \min_{k \in \mathcal{C}} \sum_{i=1}^{M_{R_k}} \cos^2 \theta_{k,\tilde{k},i},
\end{eqnarray}
which reflects the similarity of the channels between a user and the other users in the group.

Although the above-mentioned bound is not tight enough\footnote{In~\cite{Shen:07}, Shen \emph{et al.}~derived a lower bound on the ergodic sum capacity with BD in terms of unordered eigenvalues of complex Wishart matrices.  While their objective is to derive a tight semi-closed-form expression and perform an analytical comparison between BD and DPC, our derivation aims at investigating into the relationship between the sum rate capacity and the principal angles of users.}, it is sufficient to determine the relationship between one user and the other group members as will be shown by simulations in Section~V.

\subsubsection{Selection-Oriented Criterion}
While the objective of a user grouping algorithm is to serve all users once over an entire scheduling period, user selection algorithm serves only a subset of users at one time. For a selected user subset, an addition of a new user would induce a change in sum rate capacity~$\Delta {C}$ that can be separated into two components~\cite{Lee:10}, namely the incremental gain in sum rate capacity~${C}_{gain}$ and the incremental capacity degradation~${C}_{loss}$ due to the interference of this new user on the existing users.  It is clear that if the gain surpasses the loss, the incoming user would exert an positive influence on the sum rate capacity, and vice versa.  In view of this, one promising user selection criterion is to evaluate the impact of each user from the candidate user pool on the change in sum rate capacity of the user subset, followed by enrolling a user into the subset if it brings the largest and positive value of~$\Delta {C}$. In the following, we derive a user selection criterion by quantifying the capacity change in terms of the geometrical angles.

\begin{Theorem}[Change in sum rate capacity due to a new-user addition~\cite{Razi:10}]
For a $K$-heterogeneous user downlink MU-MIMO system, the change in sum rate capacity when a new user is added in a selected user subset~$\mathcal{T}$ is
\begin{eqnarray} \label{eq:C-change}
    \Delta {C} &=& {C}_{gain} - {C}_{loss},
\end{eqnarray}
where ${C}_{gain}$~quantifies the gain in sum rate capacity due to a new incoming user~(user~$k$) and it is approximated as
\begin{eqnarray} \label{eq:C-gain}
    {C}_{gain} &\approx& \log_2 \left( \frac {\rho_k}{\sigma_n^2} \det\left( \bar{\mathbf H}_k \bar{\mathbf H}_k^H  \right) \sin^2 \psi_{k,s} \right),
\end{eqnarray}
with $\psi_{k,s}=\measuredangle(\bar{\mathbf{H}}_k, {\mathbf{H}}_s)$~being the geometrical angle between the range spaces of the channel matrix of the new user~($\bar{\mathbf H}_k$) and the aggregated channel matrix of the existing users in the selected user subset~($\mathbf{H}_s$), while ${C}_{loss}$~denotes the loss in sum rate capacity resulting from the interference of this new user to existing users and it is given by
\begin{eqnarray} \label{eq:C-loss}
    {C}_{loss} &\approx& \sum_{j \in \mathcal{T}} \log_2 \left( \frac{\rho_j}{\sigma_n^2} \det\left(\bar{\mathbf H}_j \bar{\mathbf H}_j^H\right) \sin^2 \psi_{j,s\backslash j} \sin^2 \psi_{k,s\backslash j} \right),
\end{eqnarray}
with~$\psi_{j,s\backslash j}=\measuredangle(\bar{\mathbf{H}}_j, {\mathbf{H}}_s\backslash {\mathbf{H}}_j)$ being the geometrical angle between the range spaces of the channel matrix of the $j$-th~user in the subset~($\bar{\mathbf H}_j$) and the aggregated channel matrix of the other selected users~${\mathbf{H}}_s \backslash \mathbf H_j$, and similar definition holds for $\psi_{k,s\backslash j}=\measuredangle({\bar{\mathbf H}}_k, {\mathbf{H}}_s\backslash {\mathbf{H}}_j)$. $\hfill{\square}$
\end{Theorem}
\begin{Proof}
Please refer to Appendix~B for details.
\end{Proof}

Regarding the loss in sum rate capacity~${C}_{loss}$, it can be geometrically interpreted by the following example.  Suppose there are two users (user-1 and user-2) in an existing user subset and a new incoming user (user-3) whose effective channel lies in the intersection of the null spaces of~$\mathbf H_1$ and~$\mathbf H_2$. As referred to~(\ref{eq:C-loss}), the capacity loss due to user-3 is mainly due to two components: one is the projection of user-3's channel onto the null space of~$\mathbf H_1$ followed by the range space of~$\mathbf H_2$, and the other one is the projection of user-3's channel onto the null space of~$\mathbf H_2$ and then the range space of~$\mathbf H_1$. As shown in Fig.~2(b), the first component is equivalent to recursively project~$OA$ onto~$OB$ and {then~$OC$. Similarly, the second component is geometrically equivalent to a recursive projection of~$OA$ onto~$OE$ followed by~$OD$.

From Theorem~2, it is clear that the corresponding performance metric is to maximize the incremental improvement in~(\ref{eq:C-change}), i.e.,
\begin{eqnarray} \label{eq:metric-sum-rate-change}
    \mathcal{M}(\mathbf H_k, {\mathbf H}_s) &=& \arg \max_{k \in \mathcal{C}} \frac {\sigma_n^{2(|\mathcal{T}|-1)} \rho_k \det\left( \bar{\mathbf H}_k \bar{\mathbf H}_k^H \right) \sin^2 \psi_{k,s}} {\prod_{j \in \mathcal{T}} \rho_j \det\left( \bar{\mathbf H}_j \bar{\mathbf H}_j^H \right) \sin^2 \psi_{j,s\backslash j} \sin^2 \psi_{k,s\backslash j} },
\end{eqnarray}
which measures the influence introduced by the incoming user (denoted by $\mathbf H_k$) on the already selected users (denoted by ${\mathbf H}_s$ the concatenated channel matrices), and it is used as a scheduling criterion for user selection or hybrid user scheduling.

Alternatively, a simplified user selection criterion is to solely consider~${C}_{gain}$, namely
\begin{eqnarray} \label{eq:metric-sum-rate-gain}
    \mathcal{M}(\mathbf H_k, {\mathbf H}_s) &=& \arg \max_{k \in \mathcal{C}} \rho_k \det\left( \bar{\mathbf H}_k \bar{\mathbf H}_k^H \right) \sin^2 \psi_{k,s},
    \\ \label{eq:metric-sum-rate-gain-2}
    &=& \arg \max_{k \in \mathcal{C}} \rho_k \left( \det\left( \bar{\mathbf H}_k \bar{\mathbf H}_k^H \right) - \det\left( \bar{\mathbf H}_k \mathbf H_s^H \mathbf H_s \bar{\mathbf H}_k^H \right) \right)
    \\ \label{eq:metric-sum-rate-gain-3}
    &=& \arg \max_{k \in \mathcal{C}} \rho_k \det\left( \bar{\mathbf H}_k {\mathbf H}_s^{\perp H} {\mathbf H}_s^{\perp} \bar{\mathbf H}_k^H \right),
\end{eqnarray}
where~(\ref{eq:metric-sum-rate-gain-2}) is due to the alternative definition of geometrical angle as given in~(\ref{eq:user-grouping-geomatrical-angle-alter}).  Geometrically, this simplified criterion~(\ref{eq:metric-sum-rate-gain-3}) refers to the volume of the parallelepiped spanned by the projection of the basis vectors of the range space of~$\bar{\mathbf H}_k$ onto the null space of~$\mathbf H_s$, i.e., the aggregated channel matrix of the scheduled users. Moreover, when compared~(\ref{eq:metric-sum-rate-gain-3}) with an alternative expression of geometrical angle to~(\ref{eq:metric-geometrical-angle}) in terms of sine function\footnote{In~(\ref{eq:metric-geometrical-angle}), the user selection criterion is to find a user with the minimum value of the product of cosine square. It is equivalent to search for a user that maximizes the product of sine square, i.e.,~$\mathcal{M}(\mathbf H_k, {\mathbf H}_j) = \arg \max_{k,j \in \mathcal{C}} \sin^2 \psi_{k,j}$.}, their difference lies on the volume of channel matrix, i.e.,~$\rho_k \det\left( \bar{\mathbf H}_k \bar{\mathbf H}_k^H \right)$. Nevertheless, the impact of the channel volume on~$\mathcal{C}_k$ vanishes when the number of users is asymptotically large because a best user with $\rho_k \det\left( \bar{\mathbf H}_k \bar{\mathbf H}_k^H \right) \to 1$ can always be found.

To sum up, our proposed criteria can be applied in various user scheduling algorithms. In particular, the geometrical angle (c.f.,~(\ref{eq:metric-geometrical-angle})) and grouping-oriented criterion (c.f.,~(\ref{eq:metric-sum-cosine})) are more suitable for user grouping algorithms because both criteria emphasize on the integrated effect of all involved users.  On the other hand, the selection-oriented criteria (c.f.,~(\ref{eq:metric-sum-rate-change}) and~(\ref{eq:metric-sum-rate-gain})) focus more on the impact of a newly-recruited user on the sum rate capacity and therefore, they are more appropriate for user selection algorithms or our proposed hybrid user scheduling algorithms that will be presented in the next section.

\section{Proposed Scheduling Algorithms for Heterogeneous Users}
In this section, we first review a conventional user grouping algorithm that is dedicated for scheduling spatially correlated homogeneous users, followed by presenting two reduced-complexity greedy-based user grouping algorithms.

\subsection{Conventional User Grouping Algorithm~\cite{Wang:06}}
Consider a downlink MU-MIMO system with $K = 2L$~homogeneous users. Assume that the channel remains unchanged during the entire scheduling period of $L$~timeslots such that these $K$~users are divided into~$L$ groups of size~$G = 2$. The objective of~\cite{Wang:06} is to design a user scheduling algorithm so as to minimize spatial correlation between two users per group (or equivalently, in the same timeslot) while maximizing multiuser diversity. In this context, Wang and Murch~\cite{Wang:06} have considered using the largest principal angle as the user grouping criteria. In particular, given ${C_2^{2L}C_2^{2L-2} \cdots C_2^{2}}/{L !}$~possible arrangements\footnote{Each arrangement consists of~$L$ groups, each of which has $G=2$~users.}, a max-min operation is performed in which the smallest largest principal angle for each arrangement is first identified, followed by selecting the arrangement with the largest value among all these ${C_2^{2L}C_2^{2L-2} \cdots C_2^{2}}/{L !}$~smallest angles as the best one. However, there are two main drawbacks in applying this algorithm to systems with heterogeneous users.
\begin{enumerate}
\item
    \emph{Reduced Average Sum Rate Capacity.} Due to the fact that homogeneous users are equipped with the same number of receive antennas, the group size can be heuristically set as a constant~$G = M_T/M_R$, where~$M_{R_k} = M_R$ for all~$k$. For heterogeneous users, however, it is not wise to determine the group size in advance because each user may have different number of receive antennas. For example, if we set~$G = \lfloor M_T/\max{M_{R_k}}\rfloor$, it is apparent that either the total dimension/degree of freedom per group cannot be fully utilized or a larger number of groups is required\footnote{Take a~$\{ 1, 1, 1, 2, 3, 4\} \times 6$ MU-MIMO system as an example. Here, $M_T = 6$, $\max{M_{R_k}} = 4$, and $\min{M_{R_k}} = 1$. If the group size~$G$ is determined by $\lfloor M_T/\max M_{R_k}\rfloor$, then there exists only $G = \lfloor 6/4 \rfloor = 1$~user per group. In this case, there are 3~groups that separately consist of 1~user with single receive antenna, and the remaining degree of freedom (i.e., $M_T - \min{M_{R_k}} = 5$ dimensions) cannot be fully utilized.  Further, since it requires $N_G = 6$ groups (or 6~timeslots) in serving all users, it is equivalent to the conventional TDMA scheme in which only one user is served by the BS at a time.  Due to the lack of spatial multiplexing among users, the sum rate capacity per group is significantly reduced.}, which results in a lower average sum rate capacity per group. On the other hand, if we set the group size according to the minimum number of receive antennas, the total number of receive antennas in a group will definitely exceed the number of transmit antennas, which results in the violation of the dimensionality constraint of BD.
\item
    \emph{Huge Computational Complexity.} Roughly speaking, the algorithm involves as many as \\${C_2^{2L}C_2^{2L-2} \cdots C_2^{2}}/{L !}$ possible arrangements.  Since there are $L$~groups for each arrangement, more than $\mathcal{O}(L^2)$~comparisons\footnote{Here, $f(x)=\mathcal{O}(g(x))$ is used to represent $f(x)/g(x) = a$ when $x \to \infty$, where $a$~is a constant irrelative to~$x$.} are required per arrangement.
\end{enumerate}

Because of these concerns, we have developed two hybrid user scheduling algorithms that takes into account some key features of user grouping and selection algorithms, i.e., to capture fairness among users and to maximize the system performance in a greedy manner. These two algorithms, which aim at minimizing group size and maximizing degree of freedom, are outlined in Tables~I and~II, respectively, and they are summarized as follows.

\subsection{Algorithm 1: Group Number Minimization}
In contrast to the conventional user grouping algorithm that considers a constant group size, we alternatively consider variable group size and minimize the number of groups~$N_G$ required by setting
\begin{eqnarray}
    N_G &=& \left \lfloor \frac {\sum_{k=1}^K M_{R_k}} {M_T} \right \rfloor.
\end{eqnarray}
Each group is allowed to have different number of group members as long as its total dimension (i.e.,~the total number of receive antennas) is smaller than or equal to~$M_T$, i.e., the total degree of freedom available for interference-free transmission with BD.  In this case, we can ensure that the algorithm provides the same fairness as the conventional algorithm but requires a fewer number of groups\footnote{We take a $\{ 1, 1, 1, 2, 3, 4\} \times 6$ MU-MIMO system as discussed in footnote~7 as our example again.  By using this proposed algorithm, the total number of groups~$N_G$ is significantly reduced from~6 to~2.}.

For each group~$\mathcal{T}^{(g)}$, where~$g = 1, 2, \cdots, N_G$, users are selected in such a way that better users have higher priority in getting the resources. In particular, we select the best~$N_G$ users with the largest Frobenius norm from the candidate user pool~$\mathcal{C} = \{1, 2, \cdots, K\}$ and assign them to be the first user of each group, i.e.,
\begin{eqnarray}
    \left\{ \begin{array}{ll}
        u_1^{(g)} = \arg \max_{u \in \mathcal C} \|\mathbf H_u\|
        \\
        \mathcal T^{(g)} = \{ u_1^{(g)} \}, \quad \mathcal{C} = \mathcal{C} \backslash \{ u_1^{(g)} \}
    \end{array} \right..
\end{eqnarray}
For the remaining~$K - N_G$ users, they will be assigned to one of the $N_G$~groups by certain criterion that aims at minimizing subspace correlation and interference non-orthogonality per group while maximizing the sum rate capacity. Taking into account the performance-and-complexity tradeoff, we consider the simplified selection-oriented criterion as given in~(\ref{eq:metric-sum-rate-gain}). It is important mentioning that the idea of our approach is inspired by the idea of greedy selection but there are two main differences:
\begin{enumerate}
\item
    While typical user selection algorithms aims at choosing the ``best'' user for a group/user subset, our algorithm alternatively help users select the best group.
\item
    No user is allowed to be assigned into more than one group. In other words, each user is served by the BS only once within an entire scheduling period of $N_G$~timeslots.
\end{enumerate}
The selection procedure is summarized as follows.  Firstly, we identify a user (say, user-$k$) with the largest Frobenius norm from the updated candidate user pool.  Then, the simplified selection-oriented criterion is executed by selecting a group~$\mathcal{T}^{(g)}$ that has the largest incremental gain in sum rate capacity due to the enrollment of this user while the dimensionality constraint of BD is satisfied.  This procedure can also be mathematically expressed as
\begin{eqnarray}
    \left\{ \begin{array}{ll}
        u_k = \arg \max_{k \in \mathcal C} \|\mathbf H_k\|
        \\
        g_s = \arg \max_{1 \le g \le G} \rho_k \det(\bar{\mathbf H}_{k} \bar{\mathbf H}_{k}^H) \sin^2 \measuredangle (\mathbf H_s^{(g)},\mathbf H_k), \quad \mathcal{C} = \mathcal{C} \backslash \{ u_k \},
    \end{array} \right.
\end{eqnarray}
such that $rank(\mathbf H_s^{(g)}) + rank(\mathbf H_k) \le M_T$.  If the constraint cannot be satisfied, this user will be assigned to the next best group.  As an important remark, once the user is selected, there is a user shedding step~\cite{Yoo:06,Razi:10}  in which it will be removed from the candidate user pool (i.e., $\mathcal{C} = \mathcal{C} \backslash \{ u_k \}$) such that they are not being further considered in the remaining iterations.

\subsection{Algorithm~2: Degree-of-Freedom Maximization}
{The idea of this algorithm is to fully utilize the total dimension/degrees of freedom for every group which, according to the dimensionality constraint of BD, is the number of transmit antennas at the BS,~$M_T$. The algorithm is initialized by setting the group size to the number of transmit antennas, i.e.,~$G = M_T$.  User selection is started at the first group~$\mathcal{T}^{(1)}$ by choosing the first user out of the~$K$ total users in the candidate user pool~$\mathcal{C}$ with the maximum Frobenius norm. i.e.,
\begin{eqnarray}
    \left\{ \begin{array}{ll}
        u_1 = \arg \max_{u \in \mathcal C} \|\mathbf H_u\|
        \\
        \mathcal T^{(1)} = \{ u_1 \}, \quad \mathcal{C} = \mathcal{C} \backslash \{ u_1 \}
    \end{array} \right..
\end{eqnarray}
Like the previous algorithm, a user shedding step is performed such that the selected user will no longer be considered again in next iterations.

Then, the next best users~$u_k$ for~$\mathcal{T}^{(1)}$ are chosen from the updated candidate user pool according to the simplified selection-oriented criterion given in~(\ref{eq:metric-sum-rate-gain}) and it is mathematically written as
\begin{eqnarray}
    \left\{ \begin{array}{ll}
        u_k = \arg \max_{k \in \mathcal{C}} \rho_k \det(\bar{\mathbf H}_{k} \bar{\mathbf H}_{k}^H) \sin^2 \measuredangle (\mathbf H_s,\mathbf H_k)
        \\
        \mathcal T^{(1)} = \mathcal T^{(1)} \cup \{ u_k \}, \quad \mathcal{C} = \mathcal{C} \backslash \{ u_k \}
    \end{array} \right..
\end{eqnarray}
This selection process for~$\mathcal{T}^{(1)}$ is terminated when the sum of channel ranks of the existing users and the new incoming user is larger than the remaining degree of freedom, i.e.,~$rank(\mathbf H_s)+rank(\mathbf H_k) > M_T$.  Finally, the whole scheduling procedure repeats for the second group~$\mathcal{T}^{(2)}$ and so on until all of the $K$~users have been assigned.}

\subsection{Advantages}
Compared with the conventional user grouping algorithm~\cite{Wang:06}, our proposed hybrid user scheduling algorithms have the following advantages.
\begin{enumerate}
\item
    \emph{Higher Average Sum Rate Capacity.}{ The proposed algorithms provide an efficient mechanism in minimizing the number of groups required and utilize greedy-based criteria in enrolling as many users in one group as possible. In this case, the total dimension/degree of freedom per group is better exploited and the timeslots required for providing fairness for all users can be largely reduced, which results in a higher average sum rate capacity per group than~\cite{Wang:06}.}
\item
    \emph{Lower Computational Complexity.} While the conventional user grouping algorithm requires about ${L^2 C_2^{2L}C_2^{2L-2} \cdots C_2^{2}}/{L!}$~comparisons in finding an optimal grouping arrangement, these numbers are significantly required to approximately~$N_G(2L-N_G)$ and~$L(2L-1)$ for Algorithms-1 and~2, respectively.
\end{enumerate}

\section{Numerical Results}
Monte Carlo simulations are provided to evaluate the effectiveness of our three proposed user scheduling criteria (namely geometrical angle, grouping-oriented criterion and selection-oriented criterion) and the two proposed hybrid user scheduling algorithms in terms of the~10\% outage capacity~\cite{Stankovic:06}, which is defined as the rate that the channel can support with 90\%~probability.

For the sake of simplicity, we consider a general Kronecker Product Form~(KPF) channel model~\cite{Kermoal:02}  for the simulation, i.e.,
\begin{eqnarray} \label{eq:correlated-rician-channel-model}
  \mathbf H_k = \sqrt{\rho_k} \mathbf R_{r,k}^{1/2} \mathbf H_{k,w} \mathbf R_{t,k}^{1/2}
\end{eqnarray}
where $\rho_k=\frac{P_T}{M_T d_k^{\alpha}}$ is the received power of user-$k$ with $P_T$, $d_k$ and $\alpha$ being the transmit power, the distance between BS and $k$-th user, and path loss exponent, respectively.  In addition, $\mathbf{H}_{k,w} \in \mathcal C^{M_{R_k}\times M_T}$~is a zero-mean unit-variance i.i.d. complex Gaussian matrix between the BS and the $k$-th user, $\mathbf R_{r,k} \in \mathcal{C}^{M_{R_k} \times M_{R_k}}$ and~$\mathbf R_{t,k} \in \mathcal{C}^{M_T \times M_T}$ are the receive and transmit correlation matrices of user-$k$, which can be modeled as $[\mathbf R_{r,k}]_{ij} = \gamma_{r,k}^{|i-j|^2}$ and $[\mathbf R_{t,k}]_{ij} = \tau_{t,k}^{|i-j|^2}$, respectively, with correlation coefficients~$\gamma_{r,k}, \tau_{t,k}$~\footnote{This correlation model follows the Toeplitz structure in~\cite{Zelst:02,Mckay:05,Jin:07} and is valid for application scenarios where antenna elements at the receiver side are with equidistant spacings~\cite{Zelst:02}.}.

Unless stated otherwise, the simulation configurations of some key parameters are listed as follows.
\begin{itemize}
\item
    We employ BD as the linear precoding algorithm. Moreover, water-filling policy is considered for our numerical simulations even though the two theorems are derived by following an equal power allocation policy.
\item
    The number of receive antennas of the $k$-th user, $M_{R_{k}}$, is randomly chosen from $\{1,2,\cdots,N\}$ with equal probability, where $N$ is the largest number of receive antennas in the system.
\item
    As for the received power $\rho_k$, the path loss exponent~$\alpha$ is set to 3 and the distance $d_k$ is randomly generated with range $[200m, 1000m]$. Then, $\rho_k$~is normalized by the maximum possible received power when $d_k=200m$.
\item
    The receive and transmit correlation coefficients~$\gamma_{r,k}$ and~$\tau_{t,k}$ are modeled as uniformly distributed variables with range~[0,1].
\end{itemize}

As the first example, we evaluate the effectiveness of the first two proposed criteria, namely geometrical angle~(\ref{eq:metric-geometrical-angle}) and grouping-oriented criterion~(\ref{eq:metric-sum-cosine}), by comparing their performance with the largest principal angle, subspace collinearity and chordal distance. Two baseline criteria, namely exhaustive search and random selection, are also considered. Fig.~3 shows the 10\% outage capacity of these user grouping criteria for a~$\{ 1, 1, 1, 2, 3, 4\} \times 6$ MU-MIMO system. It can be observed that our proposed criteria perform well and both of them outperform the largest principal angle. Further, the grouping-oriented criterion achieves the average sum rate capacity per group of the exhaustive search method at high SNR ($\frac {P_T}{\sigma_n^2}$) while requiring less computational complexities.  In Fig.~4, we take a closer look at the difference in sum rate capacity between the largest principal angle and our proposed grouping-oriented criterion by considering different user configurations. In general, our proposed criterion performs as good as the largest principal angle in homogeneous environment (c.f.,~$\{2,2,2,2,2,2\}\times 12$ configuration) due to the fact that the latter does sufficiently reflect the spatial separability across homogeneous users. Nevertheless, the advantage of our proposed criterion becomes much more evident when the users are equipped with different number of receive antennas. In such a heterogeneous setting, it can be observed that our proposed criterion shows an improvement of 2~bps/Hz and 10~bps/Hz for $\{1,1,1,2,3,4\}\times 12$ and $\{1,2,3,4,5,6\}\times 12$ configurations, respectively, at a SNR of~40~dB. These results verify that our proposed criterion is a more appropriate performance metric in scheduling heterogeneous users in downlink MU-MIMO systems.

To illustrate the effectiveness of our third proposed metric, i.e., selection-oriented criterion, we apply it into a greedy user selection algorithm whose details are presented as follows. Let~$\mathcal C = \{ 1, 2, \cdots, K\}$ and~$\mathcal T =\emptyset$ be, respectively, the set of the candidate user pool (in which the initial state consists of all~$K$ users) and the scheduled user pool (i.e.,~a subset of selected users that will be served by the BS). The algorithm is initialized by specifying a maximum degree of freedom~$D$ available for interference-free transmission. Due to the dimensionality constraint of BD, it is usually an integer no larger than the total number of transmit antennas\footnote{If all users are equipped with single antenna, it is also equivalent to the number of users that can be supported at one scheduling timeslot.}, i.e.,~$ D = M_T$. User selection is started by choosing the first user with the maximum Frobenius norm. i.e.,~$u_1 = \arg \max_{u \in \mathcal{C}}|| \mathbf{H}_u ||_F$. The two user pools are updated accordingly as~$\mathcal T = \mathcal T \cup \{ u_1 \}$ and $\mathcal C = \mathcal C \backslash \{ u_1 \}$. Then, the next best users~$u_k$, where $k \in \mathcal{C}$, are chosen from the updated candidate user pool according to one of our proposed user selection criteria and we hereby take the simplified selection-oriented criterion given in~(\ref{eq:metric-sum-rate-gain}) as an illustrative example. The selection process will be terminated if the sum of channel ranks of the existing users in the subset and the new incoming user is larger than the remaining degree of freedom, i.e.,~$rank(\mathbf H_s)+rank(\mathbf H_k) > M_T$. Fig.~5 illustrates the effectiveness of our proposed selection-oriented criterion. We consider a heterogeneous MU-MIMO broadcast channel, where the BS has~12 transmit antennas and each of the 20~users equips with either~1 or~2 receive antennas.  For comparison purpose, we have also considered~(a) the greedy zero-forcing algorithm with BD precoding~\cite{Dimic:05,Sigdel:09} and~(b) applying the largest principal angle into the user selection algorithm.  As referred to the figure, it is clear that the largest principal angle does not perform well because of its incapability in reflecting users' spatial separability accurately.  On the other hand, our proposed simplified selection-oriented criterion performs better than the greedy zero forcing algorithm and it can be observed that the performance difference increases with SNR (for example, from less than 5~bps/Hz at 20~dB to more than 10~bps/Hz at 40~dB). In addition to the simplified selection criterion, we have also shown the performance of the original version of our proposed criterion~(\ref{eq:metric-sum-rate-change}) that considers both~${C}_{gain}$ and~${C}_{loss}$.  It is seen that the capacity improvement is even higher despite an increase in computational complexity but the performance of the simplified criterion~(\ref{eq:metric-sum-rate-gain}) would approach the original one~(\ref{eq:metric-sum-rate-change}) when the number of users is asymptotically large. It is because when the number of users in the candidate user pool increases, we can always find a user with the largest gain in sum rate capacity and a relatively smaller capacity loss.

Apart from comparing the performance of various user scheduling criteria from the SNR point of view, we also investigate into their performance in terms of the number of users available for scheduling.  In order to demonstrate the effectiveness of our proposed selection-oriented criterion~(\ref{eq:metric-sum-rate-change}), we also consider an optimal user selection by exhaustive search.  As referred to Fig.~6, it can be observed that at high SNR, the performance of our proposed selection-oriented criterion approaches that of the optimal user selection when the number of users is large enough (e.g.,~30 users). There is also an interesting observation on geometrical angle.  Namely, while its sum rate capacity is the lowest among all possible criteria when there are only a few users in the candidate user pool, its performance increases with the number of users and approaches that of our proposed selection-oriented criterion. This observation is consistent with our findings in Section~III.B that the impact of the channel volume~(i.e.,~$\rho_k \det\left( \bar{\mathbf H}_k \bar{\mathbf H}_k^H \right)$ in~(\ref{eq:metric-sum-rate-change}) and (\ref{eq:metric-sum-rate-gain})) on the sum rate capacity vanishes when the number of users is asymptotically large.

Lastly, we demonstrate the effectiveness of our proposed hybrid user scheduling algorithms over the conventional one~\cite{Wang:06} in a $\{ 1,1,1,2,3,4\} \times 6$~MU-MIMO configuration.  As discussed in~Section~IV.A, the group size of~\cite{Wang:06} is pre-determined in advance and its average sum rate capacity per group is expected to be lower than that of our proposed algorithms.  In view of these concerns, we apply an optimal user grouping strategy by exhaustive search in~\cite{Wang:06}, while considering only the sub-optimal yet simplified selection-oriented criterion~(\ref{eq:metric-sum-rate-gain}) for our two proposed algorithms. {It can be observed from the numerical results in Fig.~7 that though the performance of our two proposed algorithms are slightly inferior than~\cite{Wang:06} at low SNR due to the asymptotic SNR approximation~(\ref{eq:C-gain-3-approx}) for~${C}_{gain}$, they perform better at high SNR while requiring significantly less computational complexities. Fig.~8 also shows a relatively fair comparison in which all of the three user scheduling algorithms employ the largest principal angle as the performance metric.  As it can be seen, the Group-Number Minimized Algorithm~1 outperforms~\cite{Wang:06} for the entire SNR  of interest while the Degree-of-Freedom Maximized Algorithm~2 shows its superior performance at high SNR.  This performance difference is mainly due to the grouping arrangement in the first algorithm in which the best~$N_G$ users (in terms of the Frobenius norm) are distributed among all groups such that they are not competing with one another for resources.  These users can then be allocated resources with higher priority and exert a higher influence on the sum rate capacity. Based on these results together with the performance-and-complexity tradeoff, it is clear that our proposed hybrid user scheduling algorithms are a promising candidate to be applied in heterogeneous environment.}

\section{Conclusion}
In this paper, we have investigated into the design of user scheduling metrics for downlink MU-MIMO systems with heterogeneous users. We study users' channel characteristics in a subspace approach by representing the mutual interference across users that are originated from the interference non-orthogonality as the inter-user subspace correlation, and find that those conventional subspace-based user scheduling criteria that are commonly used in homogeneous users do not accurately reflect users' spatial separability.  In response, we design from a geometric point of view three effective user scheduling metrics that aim at maximizing sum rate capacity while minimizing interference non-orthogonality among users.  We also propose two hybrid user scheduling algorithms that can capture fairness among users while maximizing sum rate capacity in a greedy manner.  When compared with the conventional user scheduling algorithm,  our proposed approaches have lower computational complexities and shown to achieve a higher average sum rate capacity.

\appendix \renewcommand\thesubsection{Appendix \Roman{subsection}}
\subsection{Proof of Theorem 1}
For BD~\cite{Spencer:04-1}, the precoding matrix of the $k$-th user is expressed as a product of two precoders~$\mathbf{F}_{a_k}$ and~$\mathbf{F}_{b_k}$, in which the former is used for suppressing multiuser interference and the latter is used for performance optimization (e.g., sum rate maximization), i.e., \begin{eqnarray}
    \mathbf F_k = \beta \mathbf F_{a_k} \mathbf F_{b_k} = \beta \tilde{\mathbf V}_k^{(0)} \mathbf F_{b_k},
\end{eqnarray}
where the columns of~$\tilde{\mathbf V}_k^{(0)}$ act as basis vectors that span the null space of the interference channel~$\tilde{\mathbf H}_k$, and $\beta$~is chosen such that the total transmit power at the BS is less than the maximum transit power constraint $P_T$. Assuming equal power allocation, i.e.,~$\beta^2 \mathbf F_{b_k} \mathbf F_{b_k}^H = \mathbf I$, the sum rate capacity of the $k$-th user can be written as follows.
\begin{eqnarray} \label{eq:ED-line-A1}
    C_k &=& \log_2 \det{ \left( \mathbf I_{M_T} + \frac{1}{\sigma_n^2} \mathbf H_k^H \mathbf H_k \mathbf F_k \mathbf F_k^H \right) }
    \\ \label{eq:ED-line-A2}
    &=& \log_2 \det{ \left( \mathbf I_{M_T} + \frac{\rho_k}{\sigma_n^2} \bar{\mathbf H}_k^H \bar{\mathbf H}_k \mathbf F_k \mathbf F_k^H \right) }
    \\ \label{eq:ED-line-A3}
    &=& \log_2 \det{ \left( \mathbf I_{M_T} + \frac{\rho_k}{\sigma_n^2} \bar{\mathbf V}_k^{(1)} \bar{\mathbf \Sigma}_k^2 \bar{\mathbf V}_k^{(1)H} \tilde{\mathbf V}_k^{(0)} \tilde{\mathbf V}_k^{(0)H} \right) },
\end{eqnarray}
where~(\ref{eq:ED-line-A1}) is due to the zero-interference constraint of BD that ensures~$\tilde{\mathbf
H}_k \mathbf F_k = \mathbf{0}$~\cite{Spencer:04-1}, (\ref{eq:ED-line-A2}) is due to the definition of~$\mathbf{H}_k$ (i.e.,~$\mathbf H_{k} = \sqrt{\rho_k} \bar{\mathbf H}_{k}$), and (\ref{eq:ED-line-A3})~is due to the assumption of $\beta^2 \mathbf F_{b_k} \mathbf F_{b_k}^H = \mathbf I$ and the eigenvalue decomposition of~$\bar{\mathbf H}_k$, with~$\bar{\mathbf \Sigma}_k = diag\{ \lambda_{k,1},
\lambda_{k,2}, \cdots, \lambda_{k,M_{R_k}} \}$ being its singular matrix and the columns of~$\bar{\mathbf V}_k^{(1)}$ being the basis vectors spanning its range space.

Denote~$\mathbf T_k = \tilde{\mathbf V}_k^{(0)H} \bar{\mathbf V}_k^{(1)}$. The sum rate capacity can be expressed in terms of the eigenvalue matrix of~$\mathbf T_k \bar{\mathbf \Sigma}_k^2 \mathbf T_k^H$ as follows.
\begin{eqnarray} \label{eq:ED-line-A4}
    C_k &=& \log_2 \det{ \left( \mathbf I_{M_T - \tilde{L}_k} + \frac{\rho_k}{\sigma_n^2} \Lambda( \mathbf T_k \bar{\mathbf \Sigma}_k^2 \mathbf T_k^H)\right) },
\end{eqnarray}
where $\tilde{L}_k$~is the rank of~$\tilde{\mathbf{H}}_k$, and $\Lambda(\cdot)$ represents the corresponding diagonal matrix. Though (\ref{eq:ED-line-A4})~is exact, it is not easy to obtain any insight on user scheduling criteria and therefore, we resort to develop an upper and a lower bounds of~$C_k$ by using the following propositions.

\begin{Proposition}[Upper bound on the determinant of a matrix~\cite{Bucur:06}]
For any positive definite matrix~$\mathbf{M}_C$, the following relation holds
\begin{eqnarray}
    \det \left( \mathbf{M}_C \right) &\leq& \left(\frac{1}{m} tr \left( \mathbf{M}_C \right) \right)^{m},
\end{eqnarray}
where~$m$ is any positive integer. $\hfill{\square}$
\end{Proposition}

\begin{Proposition}[Trace inequality for matrix product~\cite{Komaroff:90,Lasserre:95}]
For any two Hermitian positive semi-definite matrices~$\mathbf{M}_D$ and~$\mathbf{M}_E$, there holds
\begin{eqnarray}
    \sum_{i=1}^n \lambda_i(\mathbf{M}_D) \lambda_{n-i+1}(\mathbf{M}_E) \le tr (\mathbf{M}_D \mathbf{M}_E) \le \sum_{i=1}^n \lambda_i(\mathbf{M}_D) \lambda_{i}(\mathbf{M}_E),
\end{eqnarray}
where $\lambda_i(\cdot)$ is the $i$-th singular value of the operated matrix.
$\hfill{\square}$
\end{Proposition}

By using Proposition~1, (\ref{eq:ED-line-A4})~can be upper-bounded as
\begin{eqnarray} \nonumber
    \log_2 \det{ \left( \mathbf I_{M_T - \tilde{L}_k} + \frac{\rho_k}{\sigma_n^2} \Lambda( \mathbf T_k \bar{\mathbf \Sigma}_k^2 \mathbf T_k^H) \right) } &\le& M_{R_k} \log_2 \left( 1 + \frac {\rho_k}{M_{R_k}\sigma_n^2} tr( \mathbf T_k \bar{\mathbf \Sigma}_k^2 \mathbf T_k^H) \right)
    \\ \label{eq:ED-line-A6}
    &=& M_{R_k} \log_2 \left( 1 + \frac {\rho_k}{M_{R_k}\sigma_n^2} tr( \bar{\mathbf \Sigma}_k^2 \mathbf T_k^H \mathbf T_k ) \right).
\end{eqnarray}

Denote~$\lambda_i(\mathbf T_k) = \sin \theta_{k,\tilde{k},i}$ with $\theta_{k,\tilde{k},i}$ being the $i$-th principal angle of the two subspaces $\tilde{\mathbf V}_k^{(0)}$ and $\bar{\mathbf V}_k^{(1)}$. The sum rate capacity~(\ref{eq:ED-line-A6}) can further be upper-bounded as the following closed-form expression by using Proposition~2.
\begin{eqnarray} \nonumber
    C_k &\le& M_{R_k} \log_2 \left( 1 + \frac {\rho_k}{M_{R_k}\sigma_n^2} \sum_{i=1}^{M_{R_k}} \lambda_{k,i}^2 \sin^2 \theta_{k,\tilde{k},i}\right) \\ \label{eq:ED-line-A8}
    &\le& M_{R_k} \log_2 \left( 1 + \frac {\rho_k}{M_{R_k}\sigma_n^2} \lambda_{k,max}^2 \sum_{i=1}^{M_{R_k}} \sin^2 \theta_{k,\tilde{k},i}\right),
\end{eqnarray}
where~$\lambda_{k,max} = \lambda_{k,1}$ is the maximum eigenvalue of~$\bar{\mathbf H}_k$.

Similarly, the capacity lower bound can be developed by using Proposition~2 as follows.
\begin{eqnarray} \nonumber
    C_k &=& \log_2 \det{ \left( \mathbf I_{M_T - \tilde{L}_k} + \frac{\rho_k}{\sigma_n^2} \Lambda( \mathbf T_k \bar{\mathbf \Sigma}_k^2 \mathbf T_k^H) \right) }
    \\ \nonumber
    &\ge& \log_2 \left( 1 + \frac{\rho_k}{\sigma_n^2} tr( \mathbf T_k \bar{\mathbf \Sigma}_k^2 \mathbf T_k^H)\right)
    \\ \nonumber
   &=& \log_2 \left( 1 + \frac{\rho_k}{\sigma_n^2} tr( \bar{\mathbf \Sigma}_k^2 \mathbf T_k^H \mathbf T_k)\right)
    \\ \nonumber
    &\ge& \log_2 \left( 1 + \frac{\rho_k}{\sigma_n^2} \sum_{i=1}^{M_{R_k}} \lambda_{k,n-i+1}^2 \sin^2 \theta_{k,\tilde{k},i} \right)
    \\
    &\ge& \log_2 \left( 1 + \frac{\rho_k}{\sigma_n^2} \lambda_{k,min}^2 \sum_{i=1}^{M_{R_k}} \sin^2 \theta_{k,\tilde{k},i} \right),
\end{eqnarray}
with~$\lambda_{k,mix} = \lambda_{k,M_{R_k}}$ being the minimum eigenvalue of~$\bar{\mathbf H}_k$.

This completes the proof of Theorem~1.

\subsection{Proof of Theorem 2}
For a $K$-heterogeneous user downlink MU-MIMO system, the sum rate capacity is updated by an amount~$\Delta {C}$ when a new user is added in the selected user subset~$\mathcal{T}$~\cite{Lee:10}, i.e.,
\begin{eqnarray} \nonumber
    \Delta {C} &=& {C}_{gain} - {C}_{loss},
\end{eqnarray}
where ${C}_{gain}$ refers to the gain in sum rate capacity due to this new user and it is given by
\begin{eqnarray} \nonumber
{C}_{gain} \nonumber
    &=& \log_2 \det\left( \mathbf I_{M_T} + \frac{1}{\sigma_n^2} \mathbf H_k^H \mathbf H_k \mathbf F_k \mathbf F_k^H \right)
    \\ \nonumber
    &=& \log_2 \det\left( \mathbf I_{M_T} + \frac{\rho_k}{\sigma_n^2} \bar{\mathbf V}_k \bar{\mathbf \Sigma}_k^T \bar{\mathbf \Sigma}_k \bar{\mathbf V}_k^H \mathbf V_s^{(0)} \mathbf V_s^{(0)H} \right)
    \\ \label{eq:C-gain-1}
    &=& \log_2 \det\left( \mathbf I_{M_{R_k}} + \frac{\rho_k}{\sigma_n^2}  \bar{\mathbf \Sigma}_k \bar{\mathbf \Sigma}_k^T \bar{\mathbf V}_k^{(1)H} \mathbf V_s^{(0)} \mathbf V_s^{(0)H} \bar{\mathbf V}_k^{(1)} \right),
\end{eqnarray}
with the precoder\footnote{Recall in Appendix~A that we consider the policy of equal power allocation for each precoder such that~$\beta^2 \mathbf F_{b_k} \mathbf F_{b_k}^H = \mathbf I$.}~$\mathbf{F}_k = \beta \mathbf V_s^{(0)} \mathbf{F}_{b_k}$, while the columns of~$\bar{\mathbf V}_k^{(1)}$ and~$\mathbf V_s^{(0)}$ being the basis vectors that span, respectively, the range space of the incoming user-$k$'s channel~$\bar{\mathbf{H}}_k$ and the null space of the aggregated channels of the existing users in the subset (i.e.,~$\mathbf{H}_s$).

Denote~$\mathbf \Gamma_k = \mathbf V_s^{(0)H} \bar{\mathbf V}_k^{(1)}$. We can asymptotically approximate~(\ref{eq:C-gain-1}) with respect to received SNR as
\begin{eqnarray} \nonumber
{C}_{gain}
    &=& \log_2 \det\left( \mathbf I_{M_{R_k}} + \frac{\rho_k}{\sigma_n^2} \bar{\mathbf \Sigma}_k \bar{\mathbf \Sigma}_k^T \mathbf \Gamma_k^H \mathbf \Gamma_k \right) \\ \nonumber
    &\approx& \log_2 \left( \frac{\rho_k}{\sigma_n^2} \det\left( \bar{\mathbf \Sigma}_k \bar{\mathbf \Sigma}_k^T \right) \prod_{i=1}^{M_{R_k}} \lambda_i^2({\mathbf \Gamma_k}) \right)
    \\ \label{eq:C-gain-2}
    &=& \log_2 \left(\frac{\rho_k}{\sigma_n^2} \det\left( \bar{\mathbf \Sigma}_k \bar{\mathbf \Sigma}_k^T \right) \prod_{i=1}^{M_{R_k}} \sin^2 \theta_{k,s,i}\right),
\end{eqnarray}
where~$\lambda_i(\mathbf \Gamma_k) = \sin \theta_{k,s,i}$ with $\theta_{k,s,i}$ being the $i$-th principal angle of the two subspaces~$\bar{\mathbf V}_k^{(1)}$ and~$\mathbf V_s^{(0)}$. Following the definition of geometrical angle in Section~III.B\footnote{Similarly, geometrical angle can be also represented by the sine function~\cite{Miao:96}, i.e.,~$\sin^2 \psi_{k,s} = \prod_{i=1}^p \sin^2 \theta_{k,s,i}$.}, the approximated gain in sum rate capacity~(\ref{eq:C-gain-2}) is written as
\begin{eqnarray} \label{eq:C-gain-3-approx}
    {C}_{gain} \approx \log_2 \left( \frac{\rho_k}{\sigma_n^2} \det\left( \bar{\mathbf \Sigma}_k \bar{\mathbf \Sigma}_k^T \right) \sin^2 \psi_{k,s}\right)
     = \log_2 \left(\frac {\rho_k}{\sigma_n^2} \det\left(\bar{\mathbf H}_k \bar{\mathbf H}_k^H \right) \sin^2 \psi_{k,s}\right).
\end{eqnarray}

Though the sum rate capacity is increased due to the incoming user-$k$, its presence in the subset induces interference and hence performance loss with the existing users. Denote, respectively, the sum rate capacity before and after enrolling user-$k$ as~${C}_{pre}$ and~${C}_{post}$, the loss in sum rate capacity~${C}_{loss}$ can be quantified in the following way.
\begin{eqnarray} \nonumber
    {C}_{loss} &=& {C}_{pre} - {C}_{post},
\end{eqnarray}
where
\begin{eqnarray} \nonumber
    {C}_{pre} &=& \sum_{j \in \mathcal T} \log_2 \det\left( \mathbf I_{M_T} + \frac {1}{\sigma_n^2} \mathbf H_j^H \mathbf H_j \mathbf F_j \mathbf F_j^H \right)
    \\ \label{eq:C-loss-pre}
    &=& \sum_{j \in \mathcal T} \log_2 \det\left( \mathbf I_{M_T} + \frac {1}{\sigma_n^2} \mathbf H_j^H \mathbf H_j \mathbf V_{\mathcal{T} \backslash j}^{(0)} \mathbf V_{\mathcal{T} \backslash j}^{(0)H} \right)
\end{eqnarray}
and
\begin{eqnarray} \label{eq:C-loss-post-1}
    {C}_{post} &=& \sum_{j \in \mathcal T} \log_2 \det\left( \mathbf I_{M_T} + \frac {1}{\sigma_n^2} \mathbf H_j^H \mathbf H_j \mathbf V_{(\mathcal{T} \backslash j) \cap k}^{(0)} \mathbf V_{(\mathcal{T} \backslash j) \cap k}^{(0)H} \right),
\end{eqnarray}
with~$\mathbf V_{(\mathcal{T} \backslash j) \cap k}^{(0)}$ being the intersection of the null spaces of~$\mathbf{H}_s \backslash \mathbf H_j$ and~$\mathbf H_k$. In order to make~$\mathbf V_{(\mathcal{T} \backslash j) \cap k}^{(0)}$ tractable, we apply alternating projection algorithm~\cite{Halperin:62} into~$\mathbf V_{(\mathcal{T} \backslash j) \cap k}^{(0)} \mathbf V_{(\mathcal{T} \backslash j) \cap k}^{(0)H}$ such that the intersection of two subspaces is approximated by the infinite power of the product of their projection matrices, namely,
\begin{eqnarray} \nonumber
    \mathbf V_{(\mathcal{T} \backslash j) \cap k}^{(0)} \mathbf V_{(\mathcal{T} \backslash j) \cap k}^{(0)H} &\approx&
      \left( \mathbf{V}_{T \backslash j}^{(0)} \mathbf{V}_{T \backslash j}^{(0)H} \mathbf{V}_k^{(0)} \mathbf{V}_k^{(0)H} \right)^{\kappa}
    \\
    &=& \left(\mathbf V_{\mathcal{T} \backslash j}^{(0)} \mathbf V_{\mathcal{T} \backslash j}^{(0)H} \mathbf V_{k}^{(0)}  \mathbf V_{k}^{(0)H} \mathbf V_{\mathcal{T} \backslash j}^{(0)} \mathbf V_{\mathcal{T} \backslash j}^{(0)H} \right)^{\kappa}, ~~~~ \kappa \rightarrow \infty.
\end{eqnarray}
As referred to~\cite{Fuchs:06}, Fuchs~\emph{et al.}~show by simulations that $\kappa=3$~is sufficient enough for their application scenarios of interest. Since our main focus is to investigate into the relationship of the capacity change from a geometrical viewpoint, rather than to come up with an exact closed-form expression, we consider~$\kappa=1$ for the ease of our subsequent derivation.  Then, ${C}_{post}$~in~(\ref{eq:C-loss-post-1}) can be approximated as
\begin{eqnarray} \label{eq:C-loss-post}
    {C}_{post}
    &\approx& \sum_{j \in \mathcal T} \log_2 \det\left( \mathbf I_{M_T} + \frac {1}{\sigma_n^2} \mathbf H_j^H \mathbf H_j \mathbf V_{\mathcal{T} \backslash j}^{(0)} \mathbf V_{\mathcal{T} \backslash j}^{(0)H} \mathbf V_{k}^{(0)}  \mathbf V_{k}^{(0)H} \mathbf V_{\mathcal{T} \backslash j}^{(0)} \mathbf V_{\mathcal{T} \backslash j}^{(0)H} \right),
\end{eqnarray}
with the columns of~$\mathbf V_{\mathcal{T} \backslash j}^{(0)}$ being the basis vector that span the null space of~$\mathbf{H}_s \backslash \mathbf{H}_j$, i.e., the aggregated channels of the existing users except user-$j$.

Given~(\ref{eq:C-loss-pre}) and~(\ref{eq:C-loss-post}), we have
\begin{eqnarray} \nonumber
    {C}_{loss} &=& {C}_{pre} - {C}_{post}
    \\ \nonumber
    &\approx& \sum_{j \in \mathcal T} \log_2 \det\left( \mathbf I_{M_T} + \frac {1}{\sigma_n^2} \mathbf H_j^H \mathbf H_j \mathbf V_{\mathcal{T} \backslash j}^{(0)} \mathbf V_{\mathcal{T} \backslash j}^{(0)H} \right)
    \\ \label{eq:C-loss-1}
    &-& \sum_{j \in \mathcal T} \log_2 \det\left( \mathbf I_{M_T} + \frac {1}{\sigma_n^2} \mathbf H_j^H \mathbf H_j \mathbf V_{\mathcal{T} \backslash j}^{(0)} \mathbf V_{\mathcal{T} \backslash j}^{(0)H} \mathbf V_{k}^{(0)}  \mathbf V_{k}^{(0)H} \mathbf V_{\mathcal{T} \backslash j}^{(0)} \mathbf V_{\mathcal{T} \backslash j}^{(0)H} \right).
\end{eqnarray}
Since it is well known that the following relation
\begin{eqnarray} \nonumber
    \log_2 \det(\mathbf I + \mathbf{M}_F + \mathbf{M}_G)- \log_2 \det(\mathbf I + \mathbf{M}_F) &=& \log_2 \det(\mathbf I + \mathbf{M}_G)
\end{eqnarray}
holds for any two matrices~$\mathbf{M}_F$ and~$\mathbf{M}_G$ that are orthogonal to each other (i.e.,~$\mathbf{M}_F \mathbf{M}_G^H = \mathbf 0$), (\ref{eq:C-loss-1})~is then simplified as
\begin{eqnarray} \label{eq:C-loss-2}
    {C}_{loss}
    &\approx& \sum_{j \in \mathcal T} \log_2 \det\left( \mathbf I_{M_T} + \frac {1}{\sigma_n^2} \mathbf H_j^H \mathbf H_j \mathbf V_{\mathcal{T} \backslash j}^{(0)} \mathbf V_{\mathcal{T} \backslash j}^{(0)H} \mathbf V_{k}^{(1)} \mathbf V_{k}^{(1)H} \mathbf V_{\mathcal{T} \backslash j}^{(0)} \mathbf V_{\mathcal{T} \backslash j}^{(0)H} \right).
\end{eqnarray}
Further denote $\mathbf \Upsilon_{js} = \mathbf V_{\mathcal{T} \backslash j}^{(0)H} \bar{\mathbf V}_{j}^{(1)}$ and $\mathbf \Upsilon_{ks} = \mathbf V_{\mathcal{T} \backslash j}^{(0)H} \mathbf V_{k}^{(1)}$.  We can asymptotically approximate~(\ref{eq:C-loss-2}) with respect to received SNR in the following way.
\begin{eqnarray} \nonumber
    {C}_{loss}
    &\approx& \sum_{j \in \mathcal T} \log_2 \det\left( \mathbf I_{M_{R_j}} + \frac {\rho_j}{\sigma_n^2} \bar{\mathbf \Sigma}_j \bar{\mathbf \Sigma}_j^T \mathbf \Upsilon_{js}^H \mathbf \Upsilon_{ks} \mathbf \Upsilon_{ks}^H \mathbf \Upsilon_{js} \right)
    \\ \nonumber
    &\approx& \sum_{j \in \mathcal T} \log_2 \det\left( \frac {\rho_j}{\sigma_n^2} \bar{\mathbf \Sigma}_j \bar{\mathbf \Sigma}_j^T \mathbf \Upsilon_{js}^H \mathbf \Upsilon_{ks} \mathbf \Upsilon_{ks}^H \mathbf \Upsilon_{js} \right)
    \\ \nonumber
    &=& \sum_{j \in \mathcal T} \log_2 \left( \frac{\rho_j}{\sigma_n^2} \det\left( \bar{\mathbf \Sigma}_j \bar{\mathbf \Sigma}_j^T \right) \prod_{i=1}^{M_{R_j}} \lambda^2_i(\mathbf \Upsilon_{js}) \prod_{i=1}^{M_{R_k}} \lambda^2_i(\mathbf \Upsilon_{ks}) \right)
    \\ \label{eq:C-loss-3}
    &=& \sum_{j \in \mathcal T} \log_2 \left(\frac{\rho_j}{\sigma_n^2} \det\left( \bar{\mathbf \Sigma}_j \bar{\mathbf \Sigma}_j^T \right) \prod_{i=1}^{M_{R_j}} \sin^2 \theta_{j,s\backslash j,i} \prod_{i=1}^{M_{R_k}} \sin^2 \theta_{k,s\backslash j,i}\right),
\end{eqnarray}
where~$\lambda_i(\mathbf \Upsilon_{js}) = \sin \theta_{j,s\backslash j,i}$ with $\theta_{j,s\backslash j,i}$ being the $i$-th principal angle of the two subspaces~$\mathbf V_{\mathcal{T} \backslash j}^{(0)H}$ and~$\bar{\mathbf V}_{j}^{(1)}$, and similar definition holds for~$\lambda_i(\mathbf \Upsilon_{ks})$.

Similar to the derivation of~${C}_{gain}$, we follow the definition of geometrical angle and rewrite the approximated loss in sum rate capacity as follows.
\begin{eqnarray} \nonumber
    {C}_{loss} &\approx& \sum_{j \in \mathcal T} \log_2 \left(\frac{\rho_j}{\sigma_n^2} \det\left( \bar{\mathbf \Sigma}_j \bar{\mathbf \Sigma}_j^T \right) \sin^2 \psi_{j,s\backslash j} \sin^2 \psi_{k,s\backslash j}\right)
    \\ \label{eq:C-loss-4-approx}
    &=& \sum_{j \in \mathcal T} \log_2 \left(\frac {\rho_j}{\sigma_n^2} \det\left( \bar{\mathbf{H}}_j \bar{\mathbf{H}}_j^H \right) \sin^2 \psi_{j,s\backslash j} \sin^2 \psi_{k,s\backslash j}\right).
\end{eqnarray}

With~(\ref{eq:C-gain-3-approx}) and~(\ref{eq:C-loss-4-approx}), the change in sum rate capacity due to the new incoming user~$\Delta {C}$ is obtained and expressed in terms of geometrical angles.  This completes the proof of Theorem~2.

\nocite{*} \bibliographystyle{IEEE}

\newpage
%%%-----------------------------------------Table-------------------------------------------%%
\begin{center} \textbf{Table~I} \\ \textbf{Hybrid User Scheduling Algorithm-1: Group Number Minimization}
\end{center}
\begin{algorithm}
\begin{algorithmic}[1]
\STATE $N_G=\lfloor \sum_{k=1}^K M_{R_k}/M_T \rfloor$
\STATE $\mathcal C = \{ 1, 2, \cdots, K\}$
\STATE $\mathcal T^{(1)} = \mathcal T^{(2)} = \cdots = \mathcal T^{(N_G)} = \emptyset$
\FOR{$g = 1 \rightarrow N_G$}
    \STATE {$u_1^{(g)} = \arg \max_{u \in \mathcal C} \|\mathbf H_u\|$}
    \STATE {$\mathcal T^{(g)} = \mathcal T^{(g)} \cup \{ u_1^{(g)} \}$}
    \STATE {$\mathcal C = \mathcal C \backslash \{ u_1^{(g)} \}$}
\ENDFOR
\WHILE{$|\mathcal{C}| > 0$}
    \FOR{$g = 1 \rightarrow N_G$}
        \STATE $\mathbf H_s^{(g)} = \{ \mathbf H_i, i \in \mathcal T^{(g)}\}$
    \ENDFOR
    \STATE $u_k = \arg \max_{k \in \mathcal C} \|\mathbf H_k\|$
    \STATE $g_s = \arg \max_{1 \le g \le N_G} \rho_k \det(\bar{\mathbf H}_{k} \bar{\mathbf H}_{k}^H) \sin^2 \measuredangle (\mathbf H_s^{(g)},\mathbf H_k)$    \\ s.t. $rank(\mathbf H_s^{(g)}) + rank(\mathbf H_k) \le M_T$
    \IF{$\{g_s\} \ne \emptyset$}
        \STATE $\mathcal T^{(g_s)} = \mathcal T^{(g_s)} \cup \{ u_k \}$
    \ELSE
        \STATE $N_G \leftarrow N_G+1$
        \STATE $\mathcal T^{(N_G)} = \mathcal T^{(N_G)} \cup \{ u_k \}$
    \ENDIF
    \STATE $\mathcal C = \mathcal C \backslash \{ u_k \}$
\ENDWHILE
\end{algorithmic}
\end{algorithm}

\newpage

%%%-----------------------------------------Table-------------------------------------------%%
\begin{center} \textbf{Table~II} \\ \textbf{Hybrid User Scheduling Algorithm-2: Degree-of-Fredom Maximization}
\end{center}
\begin{algorithm}
\begin{algorithmic}[1]
\STATE $\mathcal C = \{ 1, 2, \cdots, K\}$
\STATE $\mathcal T^{(1)} = \emptyset$
\STATE $g=1$
\WHILE{$|\mathcal{C}| > 0$}
    \IF{ $\mathcal T^{(g)} == \emptyset$}
        \STATE $u_1 = \arg \max_{u \in \mathcal C} \|\mathbf H_u\|$
        \STATE $\mathcal T^{(g)} = \mathcal T^{(g)} \cup \{ u_1 \}$
        \STATE $\mathcal C = \mathcal C \backslash \{ u_1 \}$
    \ELSE
        \STATE $\mathbf H_s = \{ \mathbf H_i, i \in \mathcal T^{(g)}\}$
        \STATE $u_k = \arg \max_{u \in \mathcal C} \rho_k \det(\bar{\mathbf H}_k \bar{\mathbf H}_k^H) \sin^2 \measuredangle (\mathbf H_s, \mathbf H_k)$ \\ s.t. $rank(\mathbf H_s) + rank(\mathbf H_k) \le M_T$
        \IF{$\{u_k\} \ne \emptyset$}
             \STATE $\mathcal T^{(g)} = \mathcal T^{(g)} \cup \{ u_k \}$
             \STATE $\mathcal C = \mathcal C \backslash \{ u_k \}$
        \ELSE
            \STATE $g \leftarrow g+1$
        \ENDIF
    \ENDIF
\ENDWHILE
\end{algorithmic}
\end{algorithm}

\newpage

%%%%%%%%%%%%%%%%%%%%%%%%%%%%%%%%%%%%%%%%%%%%%%%%%%%%%%%%%%%%%%%%%%%%%%%%%%%%%%%
% Figure 1 & 2
%%%%%%%%%%%%%%%%%%%%%%%%%%%%%%%%%%%%%%%%%%%%%%%%%%%%%%%%%%%%%%%%%%%%%%%%%%%%%%%
\begin{figure}[h]
\resizebox{13cm}{!}{\includegraphics{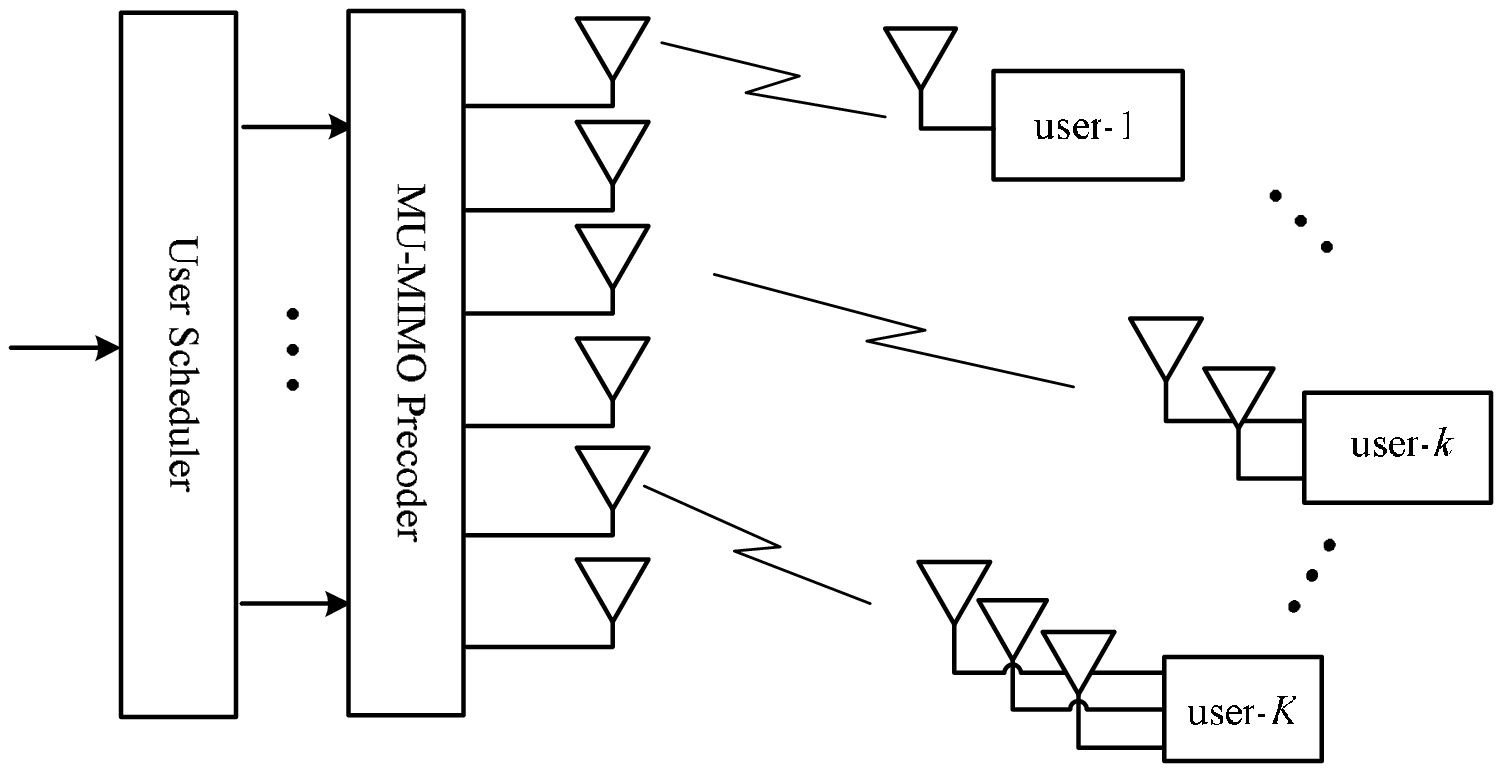}}
\vspace{-20pc}\caption{A downlink multiuser MIMO system with $M_T$ transmit antennas, and
$M_{R_k}$ receive antennas at the $k$-th user, where $k = 1, 2,
\dots, K$.}
\vspace{2pc}
\resizebox{13cm}{!}{\includegraphics{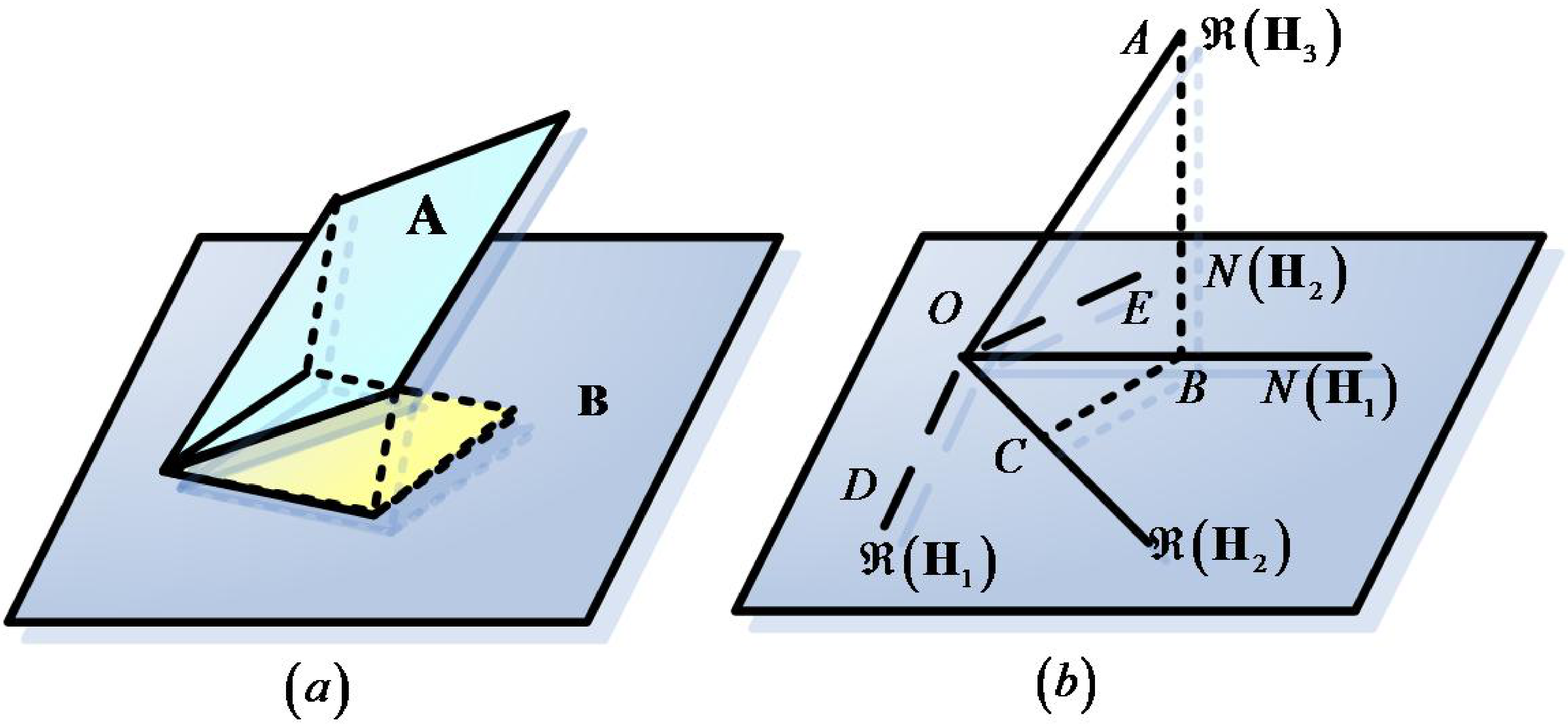}}
\caption{Geometrical illustration of (a)~geometrical angle; and (b)~the loss in sum
rate capacity due to a new incoming user.}
\end{figure}

%%%%%%%%%%%%%%%%%%%%%%%%%%%%%%%%%%%%%%%%%%%%%%%%%%%%%%%%%%%%%%%%%%%%%%%%%%%%%%%%
%% Figure 2
%%%%%%%%%%%%%%%%%%%%%%%%%%%%%%%%%%%%%%%%%%%%%%%%%%%%%%%%%%%%%%%%%%%%%%%%%%%%%%%%
%\begin{figure}
%\resizebox{13cm}{!}{\includegraphics{Fig/Geometrical-Interpret.eps}}
%\caption{Geometrical illustration of (a)~geometrical angle; and (b)~the loss in sum
%rate capacity due to a new incoming user.}
%\end{figure}

%%%%%%%%%%%%%%%%%%%%%%%%%%%%%%%%%%%%%%%%%%%%%%%%%%%%%%%%%%%%%%%%%%%%%%%%%%%%%%%
% Figure 3
%%%%%%%%%%%%%%%%%%%%%%%%%%%%%%%%%%%%%%%%%%%%%%%%%%%%%%%%%%%%%%%%%%%%%%%%%%%%%%%
\begin{figure}[h]
\begin{center}
\includegraphics[width=0.63\columnwidth]{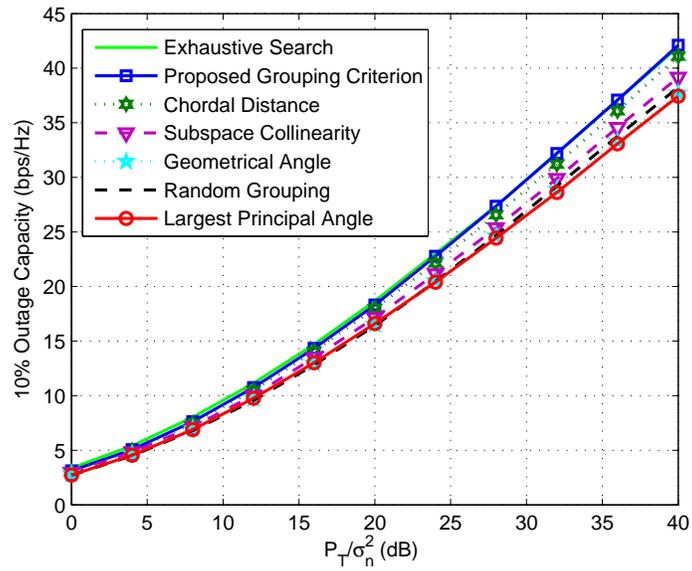}
\caption{10\% outage capacity performance of various user grouping criteria. A
$\{1,1,1,2,3,4\} \times 6$ MU-MIMO system is considered.  The
proposed criterion refers to the ``Grouping-Oriented Criterion''
in Section~III.B.2.}
\end{center}
\end{figure}

%%%%%%%%%%%%%%%%%%%%%%%%%%%%%%%%%%%%%%%%%%%%%%%%%%%%%%%%%%%%%%%%%%%%%%%%%%%%%%%
% Figure 4
%%%%%%%%%%%%%%%%%%%%%%%%%%%%%%%%%%%%%%%%%%%%%%%%%%%%%%%%%%%%%%%%%%%%%%%%%%%%%%%
\begin{figure}[h]
\begin{center}
\includegraphics[width=0.63\columnwidth]{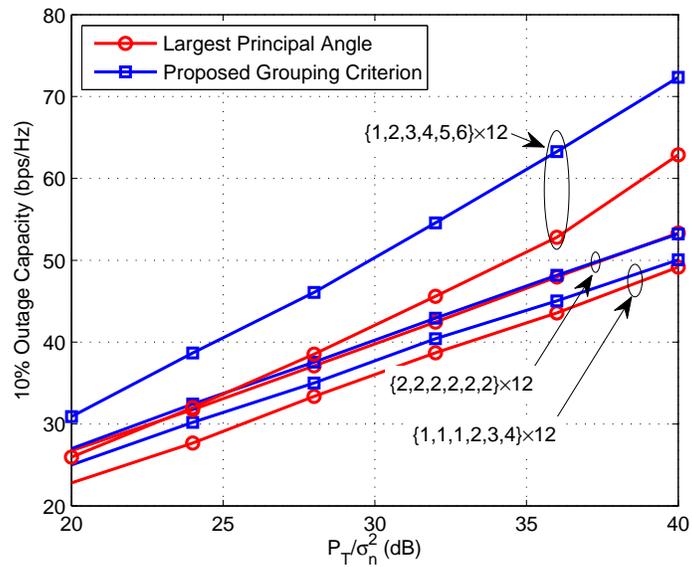}
\caption{10\% outage capacity performance of three different MU-MIMO
configurations, namely ~$\{1,2,3,4,4,4\} \times 12$,
$\{2,2,2,2,2,2\} \times 12$,~and $\{1,2,3,4,5,6\} \times 12$. The two user grouping
criteria considered are the largest principal angle and the proposed
``Grouping-Oriented Criterion'' in Section~III.B.2.}
\end{center}
\end{figure}

%%%%%%%%%%%%%%%%%%%%%%%%%%%%%%%%%%%%%%%%%%%%%%%%%%%%%%%%%%%%%%%%%%%%%%%%%%%%%%%
% Figure 5
%%%%%%%%%%%%%%%%%%%%%%%%%%%%%%%%%%%%%%%%%%%%%%%%%%%%%%%%%%%%%%%%%%%%%%%%%%%%%%%
\begin{figure}
\begin{center}
\includegraphics[width=0.63\columnwidth]{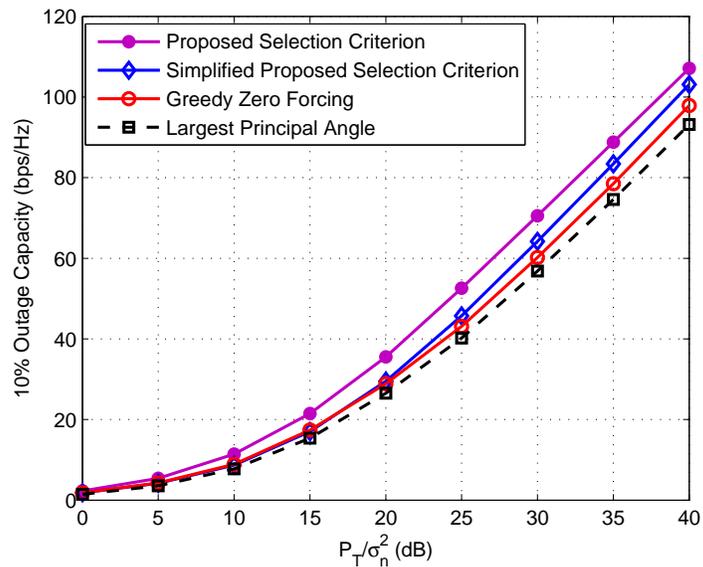}
\caption{10\% outage capacity performance of various user selection criteria.
The MU-MIMO system configuration consists of $M_T=12$ transmit
antennas at the base station and 20~users with either~1 or~2 receive
antennas.  The proposed criterion refers to the ``Selection-Oriented
Criterion'' in Section~III.B.3.}
\end{center}
\end{figure}

%%%%%%%%%%%%%%%%%%%%%%%%%%%%%%%%%%%%%%%%%%%%%%%%%%%%%%%%%%%%%%%%%%%%%%%%%%%%%%%
% Figure 6
%%%%%%%%%%%%%%%%%%%%%%%%%%%%%%%%%%%%%%%%%%%%%%%%%%%%%%%%%%%%%%%%%%%%%%%%%%%%%%%
\begin{figure}[h]
\begin{center}
\includegraphics[width=0.63\columnwidth]{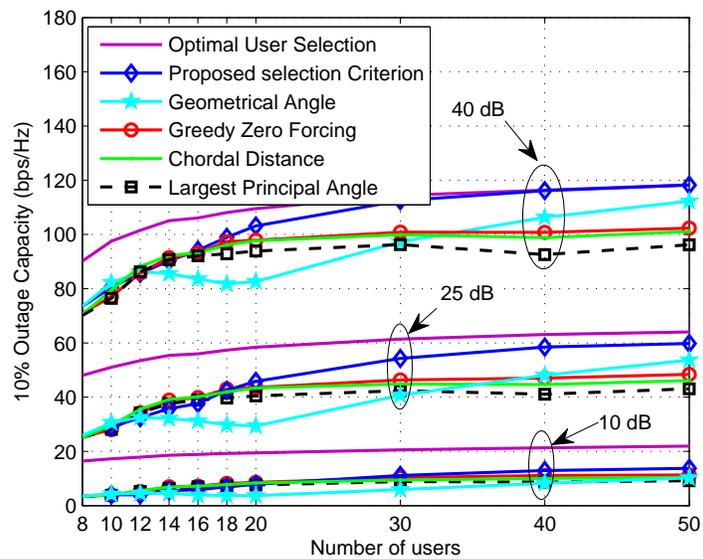}
\caption{The impact of the number of users on the 10\% outage capacity
performance of various user selection criteria. The MU-MIMO system
configuration consists of $M_T=12$ transmit antennas at the base
station.  Each user is equipped with either~1 or~2 receive antennas. The
proposed criterion refers to the ``Selection-Oriented Criterion''
in Section~III.B.3.}
\end{center}
\end{figure}

%%%%%%%%%%%%%%%%%%%%%%%%%%%%%%%%%%%%%%%%%%%%%%%%%%%%%%%%%%%%%%%%%%%%%%%%%%%%%%%
% Figure 7
%%%%%%%%%%%%%%%%%%%%%%%%%%%%%%%%%%%%%%%%%%%%%%%%%%%%%%%%%%%%%%%%%%%%%%%%%%%%%%%
\begin{figure}[h]
\begin{center}
\includegraphics[width=0.63\columnwidth]{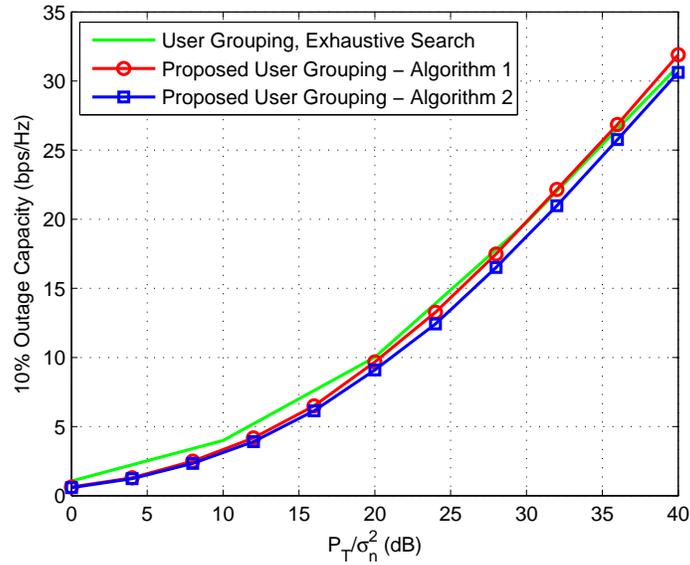}
\caption{10\% outage capacity performance of the two proposed hybrid user
scheduling algorithms (Group Number Minimized Algorithm~1 and Degree-of-Freedom Maximized Algorithm~2) and the user grouping algorithm~\cite{Wang:06}.
A $\{1,1,1,2,3,4\} \times 6$~MU-MIMO system is considered.  ``Selection-Oriented Criterion'' as presented in
Section~III.B.3 is applied for the proposed algorithms while exhaustive search
is applied for~\cite{Wang:06}.}
\end{center}
\end{figure}

%%%%%%%%%%%%%%%%%%%%%%%%%%%%%%%%%%%%%%%%%%%%%%%%%%%%%%%%%%%%%%%%%%%%%%%%%%%%%%%
% Figure 8
%%%%%%%%%%%%%%%%%%%%%%%%%%%%%%%%%%%%%%%%%%%%%%%%%%%%%%%%%%%%%%%%%%%%%%%%%%%%%%%
\begin{figure}[h]
\begin{center}
\includegraphics[width=0.63\columnwidth]{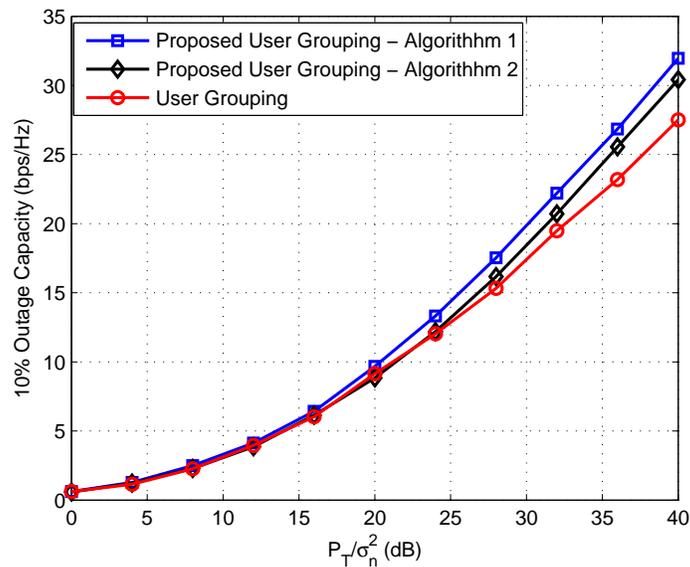}
\caption{10\% outage capacity performance of the two proposed hybrid user
scheduling algorithms (Group Number Minimized Algorithm~1 and Degree-of-Freedom Maximized Algorithm~2)  and the user grouping algorithm~\cite{Wang:06}.
A $\{1,1,1,2,3,4\} \times 6$~MU-MIMO system is considered.  All user scheduling algorithms
apply the largest principal angle as a user scheduling metric.}
\end{center}
\end{figure}

\end{document}